\documentclass[11pt,english,openany,a4]{article}
\usepackage{a4wide,amsmath,bm,booktabs,array,parskip}
\usepackage{natbib}
\usepackage[english]{babel}
\usepackage{blindtext}
\usepackage{multirow}
\usepackage{verbatim}
\usepackage{appendix}
\usepackage{hyperref} 
\usepackage{graphicx}
\usepackage{float}   
\usepackage{footmisc}
\usepackage{amsfonts}
\usepackage{amssymb}
\usepackage{bbm}
\usepackage{xcolor}
\makeatletter
\renewcommand{\paragraph}{\@startsection{paragraph}{4}{0ex}%
   {-3.25ex plus -1ex minus -0.2ex}%
   {1.5ex plus 0.2ex}%
   {\normalfont\normalsize\bfseries}}
\makeatother

\stepcounter{secnumdepth}
\stepcounter{tocdepth}
\usepackage{authblk}

\let\oldabstract\abstract
\makeatletter
\renewcommand\abstract{%
  \providecommand\keywords{\par\medskip\noindent\textit{Keywords:}\xspace}
  \oldabstract\noindent\ignorespaces}
\makeatother

\usepackage{siunitx}

\usepackage[all]{onlyamsmath}

\usepackage{natbib}
\usepackage{algorithmic, algorithm}
\usepackage{subfig}
\makeatletter
\DeclareRobustCommand*\subref{\@ifstar\sf@@subref\sf@subref}   
\makeatother

\begin{document}

\title{Modelling aggregation on the large scale and regularity on
  the small scale in spatial point pattern datasets}

\author[1,2]{Fr\'{e}d\'{e}ric Lavancier}
\author[3]{Jesper M{\o}ller}
\affil[1]{Laboratoire de Math\'{e}matiques Jean Leray\\
University of Nantes\\ Frederic.Lavancier@univ-nantes.fr}
\affil[2]{Inria\\ Centre Rennes  Bretagne Atlantique}
\affil[3]{Department of Mathematical Sciences\\
Aalborg University\\ jm@math.aau.dk}

\date{}

\maketitle

\begin{abstract} We consider a dependent
  thinning of a regular point process with the aim of obtaining
aggregation on the large scale and regularity on
  the small scale in the resulting target point process of retained
  points. Various parametric models for the
  underlying processes are
  suggested and the properties of the target point process are
  studied. Simulation and inference procedures are discussed 
  when a realization of the target point process
  is observed, depending on whether the thinned points are
  observed or not. The paper extends previous work by Dietrich Stoyan
  on interrupted point processes.

\end{abstract}

\section{Introduction}

In the spatial point process literature, analysis of
spatial point pattern datasets are often classified into
three main cases (see e.g.\ \cite{cressie:93},
\cite{Diggle:03}, and
\cite{moeller:waagepetersen:00}):
\begin{enumerate}
\item[(i)] Regularity (or inhibition or repulsiveness)---modelled by
Gibbs point processes 
\citep{ruelle:69,lieshout:00,chiu:stoyan:kendall:mecke:13}, 
Mat{\'e}rn hard core models of types I-III 
\citep{matern:86,moeller:huber:wolpert:10}, other types of hard core
processes \citep{illian:penttinen:stoyan:stoyan:08},
and determinantal point processes
\citep{Macchi:75,LMR15}.
\item[(ii)] Complete spatial randomness---modelled by Poisson
point processes \citep{kingman:93}.
\item[(iii)] Aggregation (or clustering)---modelled by Poisson cluster
  processes \citep{daley:vere-jones:03}, Cox processes \citep{cox:72}, and 
permanental point processes \citep{Macchi:75,McCullagh:Moeller:06}.
\end{enumerate}
A popular and simplistic way to determine (i)-(iii) 
is in terms of the so-called 
pair
correlation function \citep{illian:penttinen:stoyan:stoyan:08}:
Denote $\rho$ and $g$ the intensity and pair
correlation functions for a spatial point process defined on the
$d$-dimensional Euclidean space $\mathbb R^d$ (with $d=2$ or $d=3$
in most applications; formal definitions of $\rho$ and $g$
are given in Section~\ref{s:2.1}). 
For ease of presentation, we assume second order
stationarity and isotropy, i.e.\ $\rho$ is constant and for any
locations $x_1,x_2\in\mathbb R^d$,
$g(x_1,x_2)=g_0(r)$ depends only on the distance $r=\|x_1-x_2\|$. 
Intuitively, $\rho\,\mathrm dx$ is
the probability that the process has a point in an infinitesimally small
region of volume (Lebesgue measure) $\mathrm dx$,
and for $r=\|x_1-x_2\|>0$,
$g_0(r)\rho^2\,\mathrm dx_1\,\mathrm dx_2$
is the probability that the process
 has a point in each of an infinitesimally small
region around $x_1,x_2$ of volumes $\mathrm dx_1,\,\mathrm dx_2$.
Typically, $g_0(r)$ tends to 1 as $r\rightarrow\infty$, and
we are usually interested in the behaviour of $g_0(r)$ for small and
modest values of $r$. We expect
in case of (i), $g_0(r)<1$ when $r$ is small, and $g_0(r)$ is  less
than or fluctuating around 1
otherwise; in case
of (ii), $g_0=1$; and typically, in case
of (iii), 
$g_0>1$. 

For applications the classification (i)-(iii) can be too simplistic, 
and there is a lack of useful spatial point
process models with, loosely speaking,
aggregation on the large scale and regularity on
the small scale. One suggestion of such a model 
is a dependent thinning of e.g.\ a Poisson cluster point process where 
the thinning is similar to that in a
Mat{\'e}rn hard core process 
(see \cite{andersen:hahn:15}) 
or to that in
spatial birth-death constructions for Gibbs point processes (see 
\cite{kendall00} and \cite{moeller:waagepetersen:00}). 
Theoretical expressions for intensity and pair correlation of such 
Mat{\'e}rn
thinned point processes have been derived by Palm theory, and their numerical evaluation can be obtained by approximations, cf.\  
\cite{andersen:hahn:15}, while the spatial birth-death constructions 
are mathematical intractable.
Another pos\-si\-bi\-li\-ty is to consider a Gibbs
point process with a well-chosen potential that incorporates
inhibition at small scales and attraction at large scales. A famous
example is the Lennard-Jones pair-potential
\citep{ruelle:69}, and other specific potentials of this type can be
found in \cite{goldstein2014}. Unfortunately, in general for Gibbs
point processes the intensity and the
pair correlation function are unknown and 
simulation requires elaborate MCMC methods \citep{moeller:waagepetersen:00}.

This paper discusses instead a model for
 a spatial point process $X$ 
obtained by an independent thinning of a 
spatial point process $Y$ 
where the selection probabilities are given by a random process
$\Pi=\{\Pi(x):x\in\mathbb R^d\}\subseteq[0,1]$: We 
view $X$ and
$Y$ as random locally finite subsets of $\mathbb R^d$, and let
\begin{equation}\label{e:th}
X=\{x\in Y:\Pi(x)\geq U(x)\}
\end{equation}
where 
$U=\{U(x):x\in\mathbb R^d\}$ consists of independent
uniformly distributed random variables between 0 and 1, and where
$Y,\Pi,U$
are mutually independent. Dietrich Stoyan 
\citep{stoyan:79,chiu:stoyan:kendall:mecke:13}
called $X$ an interrupted
point process, which we agree is a good terminology when each $\Pi(x)$ is
either 0 or 1; indeed, in all Stoyan's examples of applications, this is
the case, though the general theory presented is not making this
restriction.   
Clearly, $\Pi$
should not be deterministic, because then the pair correlation
functions for $X$ and $Y$ would be identical ($g_X=g_Y$).
We have in mind that a realization of $X$ is observed within a bounded
window $W$, while we treat $(\Pi,U)$ as being unobserved, and 
$Y$ as being or not being
observed within $W$. 
For instance, we can think of $Y$ as describing an inhibitive behaviour
of some plant locations under optimal conditions, and $X$ as the actual 
plant locations due to unobserved covariates (e.g.\ light conditions, level of
water underground, and quality of
soil). 

\begin{figure}[h]
\begin{center}
\begin{tabular}{cc} 
\includegraphics[scale=0.25]{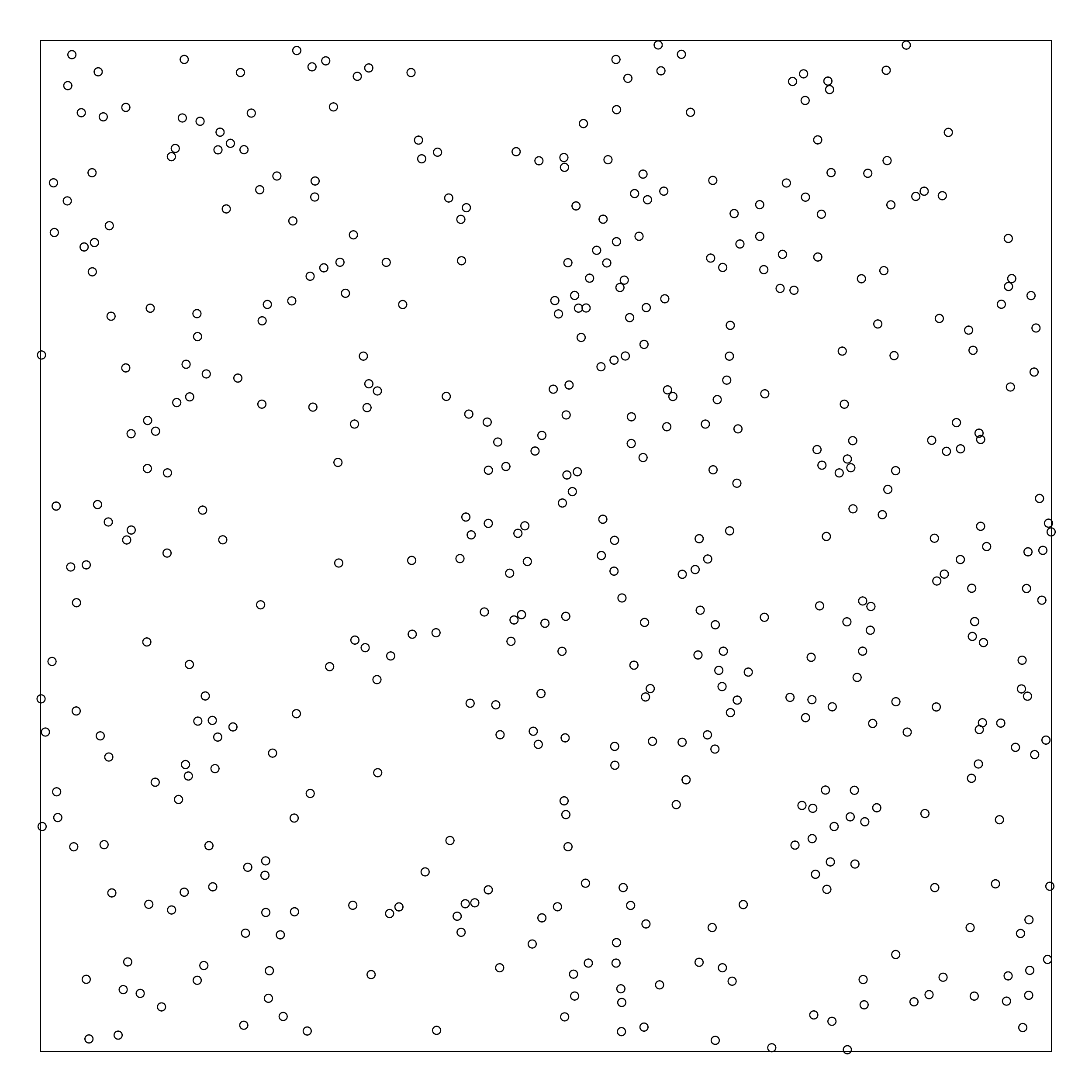} &  \includegraphics[scale=0.25]{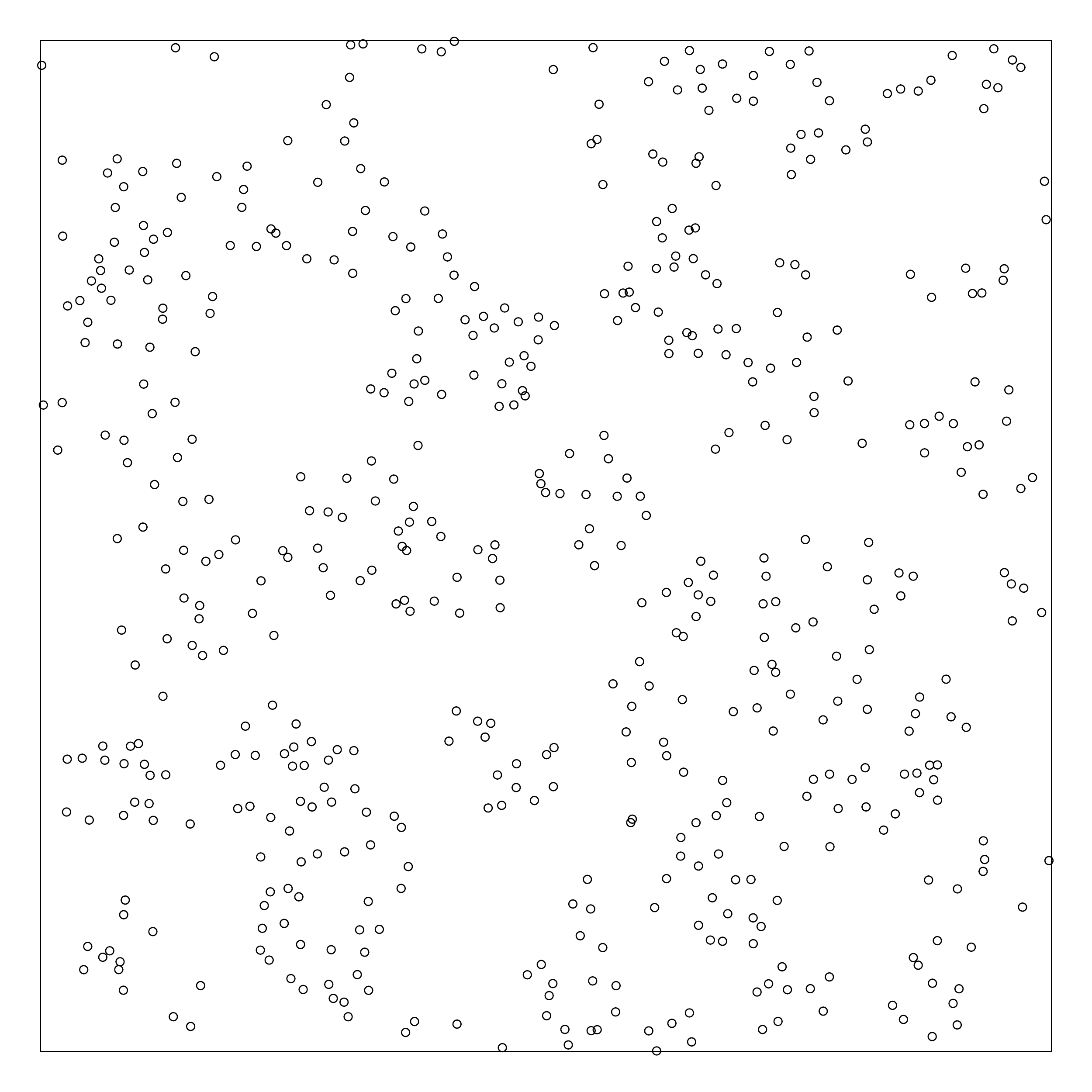}\\
 \includegraphics[scale=0.25]{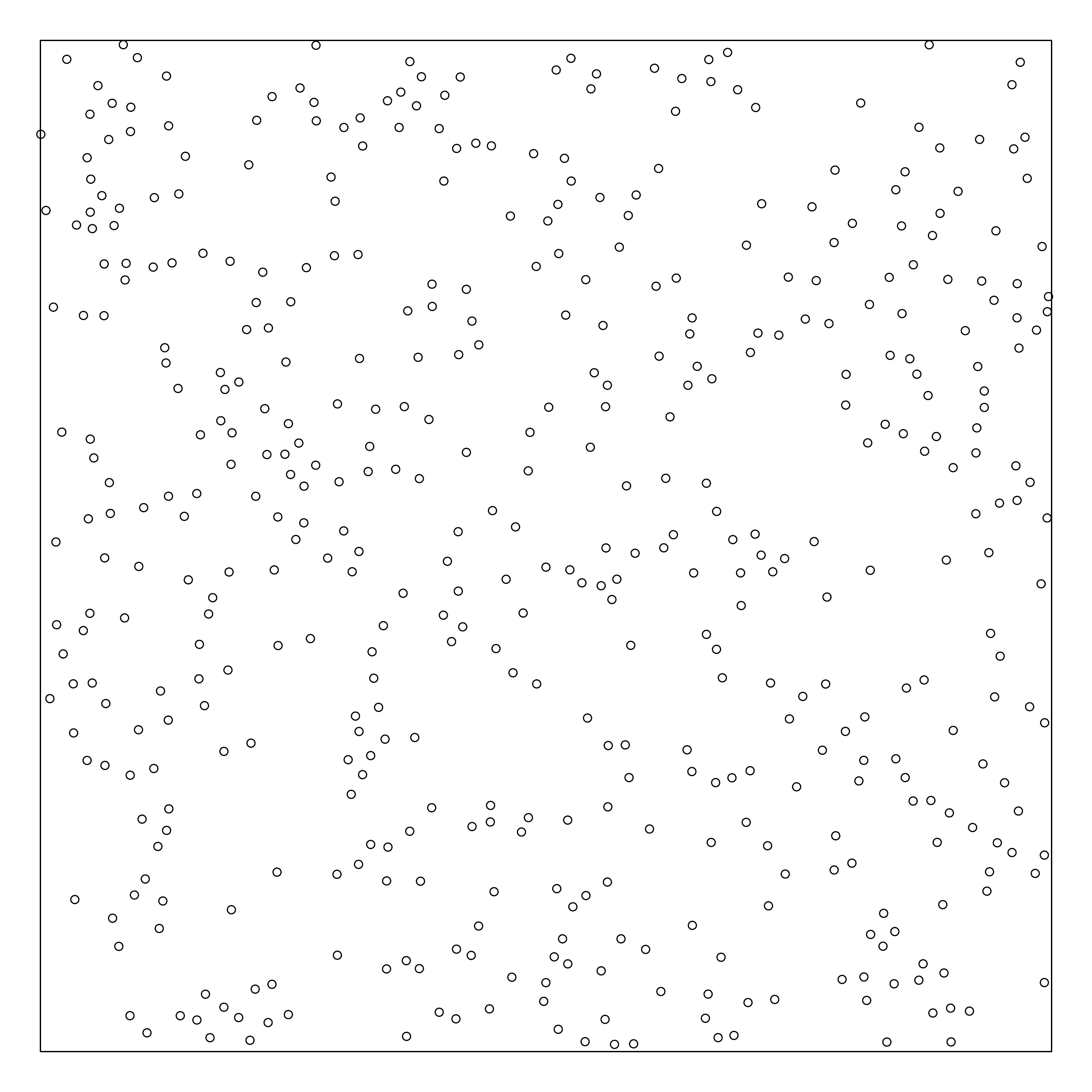}  & \includegraphics[scale=0.25]{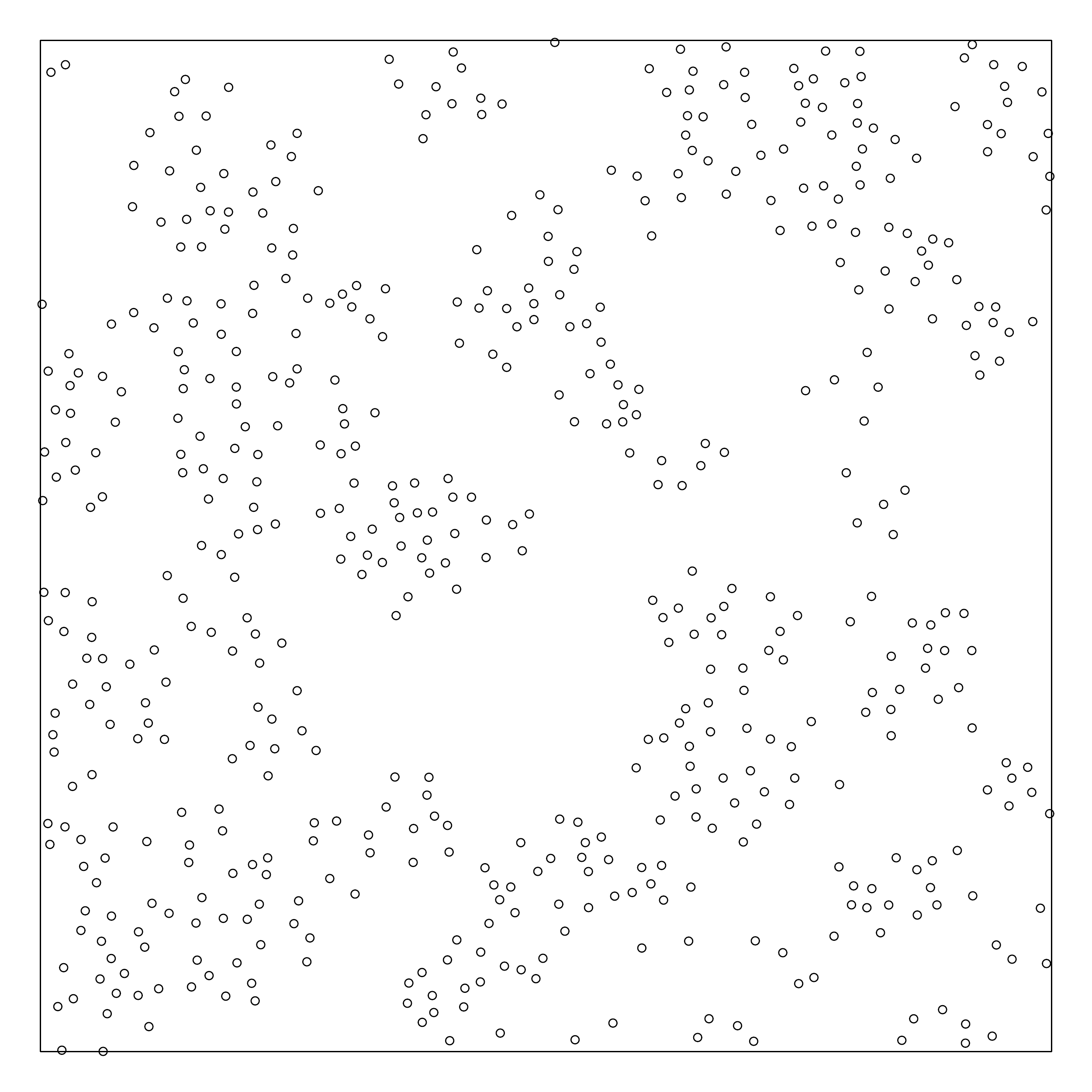} 
\end{tabular}
\caption{Examples of realizations within a unit square of $X$ 
  given by the models~1.-4.\ in Section~\ref{s:simu}: In the first row
  $Y$ is a determinantal point process, in the second row $Y$ is
 a Mat{\'e}rn
hard core process of type II, in the first column 
$-\log\Pi$ is a $\chi^2$-process, and in the second column $\Pi$
is the characteristic function for a Boolean disc model.}\label{fig:examples}
\end{center}
\end{figure}

Our idea is that it is possible to choose models for $Y$ from the class
(i) above together with models for
$\Pi$ such that $X$ exhibits small scale
regularity and large scale aggregation. 
Some examples of simulated realizations of this kind of models
 are shown in Figure~\ref{fig:examples}.
Our idea is 
demonstrated in  Sections~\ref{s:3}-\ref{s:applications}
and it can be briefly understood as follows.
Section~\ref{s:2.1} establishes a simple relationship between  
the pair correlation functions for $X$ and $Y$: 
For simplicity, assume second order stationarity
and isotropy of both $Y$ and $\Pi$, whereby our target point process
$X$ becomes second order stationary
and isotropic. The
(isotropic) pair correlation function of $X$ is given by
\begin{equation}\label{e:pcfs}
g_{X,0}(r)=M_0(r)g_{Y,0}(r)
\end{equation}
where, setting $0/0=0$, 
\begin{equation}\label{e:star}
M_0(r)=M(x_1,x_2)=\frac{\mathrm E[\Pi(x_1)\Pi(x_2)]}
{\mathrm E[\Pi(x_1)]\mathrm E[\Pi(x_2)]},\quad x_1,x_2\in\mathbb
R^d,\end{equation} 
depends only on $r=\|x_1-x_2\|$. 
For example, if $\Pi$ is positively correlated
(i.e.\ $M_0>1$) and $Y$ is a determinantal point process (this process is
described in Section~\ref{s:2.2}), then $g_{Y,0}\le 1$ and
in accordance with
\eqref{e:pcfs}
we may obtain a behaviour of $g_{X,0}$ as we wish, namely 
that $g_{X,0}$ is smaller respective larger than
1 on the small respective large scale. Examples appear later in
Figure~\ref{fig:pcf-gauss}. 

We thank Ute Hahn for reminding us about Stoyan's interrupted point
process in \cite{stoyan:79}, \cite{chiu:stoyan:kendall:mecke:13}, and \cite{kautz:et:al:11}.
In \cite{stoyan:79} and \cite{chiu:stoyan:kendall:mecke:13} he considered  
mainly the planar case where  $Y$ is a Mat{\'e}rn
hard core process of type II and where $\Pi$ is the characteristic
function of a motion invariant random closed set whose distribution
apart from $\mathrm E[\Pi(o)]$ and $M_0(r)$ is unspecified ($o$
denotes the origin in $\mathbb R^d$). In contrast we consider various
 models for both $Y$ and $\Pi$, where e.g.\ our
$\chi^2$-process model for $-\log\Pi$ seems more realistic for
applications like the example studied in Section~\ref{s:allogny}. Moreover, we
discuss simulation and parametric inference procedures depending on
whether $Y$ is observed or is a latent process. Finally, we notice
that the 
paper \cite{kautz:et:al:11} is in another direction than ours, since 
they consider $Y$ to be a
Mat{\'e}rn cluster process (which is of class (iii) above) and $\Pi$
to be the characteristic function for a motion invariant
random closed set, i.e.\ 
$X$ becomes of class (iii).

Our paper is organized as follows. 
Section~\ref{s:2} recalls some background material and deals with some
inhibitive point process models for $Y$ where $\rho_Y$ and $g_{Y,0}$
are known, namely
determinantal point processes and
Mat{\'e}rn hard core models of type I or II. 
Section~\ref{s:3} introduces models for $\Pi$, based on transformed
Gaussian processes and Boolean models, 
which combined with the models for $Y$ allow us to further study the 
behaviour of
$g_{X,0}$. Section~\ref{s:inference} discusses first simulation of $(Y,\Pi,X)$,
and second
inference for parametric models for $Y$ and $\Pi$, depending on
whether $Y$ is observed or not. Finally, Section~\ref{s:applications} 
fits parametric  models to examples of spatial point pattern datasets 
using the methodology from Section~\ref{s:inference}.

\section{Preliminaries}\label{s:2}

Let the situation be as in Section~1.
This section recalls the definitions and some properties of
product densities for a spatial point process in general and for 
determinantal point processes and
Mat{\'e}rn hard core models of types I-II in particular.

\subsection{Product densities and assumptions}\label{s:2.1}

For $n=1,2,\ldots$, suppose that $\rho^{(n)}_{Y}:\mathbb R^{dn}\mapsto[0,\infty)$ is
a Borel function satisfying the so-called Campbell formula
\begin{equation}\label{e:defrhon}
\mathrm E\sum_{x_1,\ldots,x_n\in Y}^{\not=}f(x_1,\ldots,x_n)=
\int\cdots\int f(x_1,\ldots,x_n)\rho_{Y}^{(n)}(x_1,\ldots,x_n)
\,\mathrm dx_1\cdots\,\mathrm dx_n
\end{equation}
for any non-negative Borel function $f$, where $\sum^{\not=}$ means
that $x_1,\ldots,x_n$ are pairwise distinct. Then $\rho^{(n)}_{Y}$ is
called an 
$n$th order product density of
$Y$. Such a function is 
apart from a Lebesgue nullset uniquely determined by the Campbell
formula. Henceforth, for ease of
presentation, we ignore 
nullsets. In particular, 
$\rho_{Y}=\rho_{Y}^{(1)}$ is the intensity function.
Furthermore,
setting $0/0=0$, the
pair correlation function is defined by
$g_{Y}(x_1,x_2)={\rho_{Y}^{(2)}(x_1,x_2)}/[{\rho_{Y}(x_1)\rho_{Y}(x_2)}]$.
The usual practice is to set
$\rho^{(n)}_{Y}(x_1,\ldots,x_n)=0$ if $x_i=x_j$ for some
$i\not=j$. An exception is the case of a Poisson process $Y$
where often one takes
$\rho_{Y}^{(2)}(x_1,x_2)=\rho_{Y}(x_1)\rho_{Y}(x_2)$
so that $g_{Y}(x_1,x_2)=1$ if $\rho_{Y}(x_1)\rho_{Y}(x_2)>0$. 

Recall that we assume for simplicity that $Y$
 is second order stationary and isotropic. We also
assume that
the first and second order intensity functions exist. Thus we can consider
the versions where $\rho_{Y}$ is constant and where 
$g_{Y}(x_1,x_2)=g_{Y,0}(r)$
depends only on the distance $r=\|x_1-x_2\|$. 
We call $g_0$ 
the isotropic pair correlation function.  
We furthermore assume $\rho_Y>0$  
(otherwise $Y$ is empty, which is not a case of interest). 

Similarly, $\rho^{(n)}_{X}$ and $g_{X,0}$ denote the 
$n$th order product density and the isotropic pair correlation function of
$X$. By conditioning on $Y$ and
using~\eqref{e:th} and \eqref{e:defrhon} it is straightforwardly
verified that   $\rho^{(n)}_X$ exists whenever
$\rho^{(n)}_Y$ exists, in which case
\[ 
\rho^{(n)}_X(x_1,\ldots,x_n)=
\mathrm E[\Pi(x_1)\cdots\Pi(x_n)]\rho^{(n)}_Y(x_1,\ldots,x_n).
\]
Consequently, for any $r\ge0$, 
\begin{equation}\label{e:rhonrhon}
\rho_X=q\rho_Y,\quad g_{X,0}(r)=M_0(r)g_{Y,0}(r),
\end{equation}
where $q=\mathrm E[\Pi(o)]$ denotes the mean selection probability. 
Equation \eqref{e:rhonrhon} is similar to results given in \cite{stoyan:79}.

For later purposes, denote $\bar X=Y\setminus X$ the complementary set of $X$ in $Y$. 
Then, using an obvious notation, 
\[\rho^{(n)}_{\bar X}(x_1,\ldots,x_n)=
\mathrm
E[(1-\Pi(x_1))\cdots(1-\Pi(x_n))]\rho^{(n)}_Y(x_1,\ldots,x_n)\]
 and hence by \eqref{e:star} and \eqref{e:rhonrhon},
\begin{equation}\label{e:rhonrhon_comp}
\rho_{\bar X}=(1-q)\rho_Y,\quad g_{\bar X,0}(r)=\frac{1-2q+q^2M_0(r)}{(1-q)^2}\,g_{Y,0}(r).
\end{equation}
Finally, the cross pair
correlation between $X$ and $\bar X$ (see e.g.\ 
\cite{moeller:waagepetersen:00} for a definition) is given by
\begin{equation}\label{e:rhonrhon_cross}
g_{X\bar X,0}(r)= \frac{q-q^2 M_0(r)}{q(1-q)}\,g_{Y,0}(r).
\end{equation}

\subsection{Determinantal point processes}\label{s:2.2}

Let $C$ be a complex function defined on $\mathbb R^d\times\mathbb
R^d$ and $Y$ be a determinantal point process (DPP) with kernel
$C$. By definition this means that for any
$n=1,2,\ldots$ and any $x_1,\ldots,x_n\in\mathbb R^d$,
$\rho^{(n)}_Y(x_1,\ldots,x_n)$ exists and is equal to the determinant of the
$n\times n$ matrix with $(i,j)$th entry $C(x_i,x_j)$. For background
material on DPPs, including conditions for their existence,
see \cite{LMR15} and the references
therein.

For simplicity and specificity, we assume that
$C$ is a stationary and isotropic
covariance function, i.e.\ $C(x,y)=C_0(\|x-y\|)$ is real and
non-negative definite. 
Clearly, $Y$ is then a stationary and isotropic DPP, 
and we write $Y\sim{\mathrm{DPP}}(C_0)$. 
We also assume that $C_0(\|x\|)$ 
is continuous and square integrable,
i.e.\ $\int_0^\infty r^{d-1}|C_0(r)|^2\,\mathrm dr<\infty$. 
By Theorem 2.3 and Proposition~3.1 in \cite{LMR15},
the existence of ${\mathrm{DPP}}(C_0)$ 
is then equivalent to that $\varphi_0\le1$, where
\[\varphi_0(r)=
\left\{
\begin{array}{ll}
2\int_0^\infty C_0(s)\cos(2\pi rs)\,\mathrm ds & \mbox{if $d=1$}\\
\frac{2\pi}{r^{d/2-1}}\int_0^\infty C_0(s)
J_{d/2-1}(2\pi rs)s^{d/2}\,\mathrm ds & \mbox{if $d=2,3,\ldots$}
\end{array}
\right.
\]
is the spectral density associated to $C_0$ and 
$J_{d/2-1}$ is the Bessel function of order $d/2-1$.

Observe that 
\begin{equation}\label{e:rhoDPP}
\rho_Y=C_0(0)
\end{equation} 
which is assumed to be strictly positive, and
\begin{equation}\label{e:dpppcf}
g_{Y,0}(r)=1-
{C_0(r)^2}/{\rho_{Y}^2}
\end{equation}
is 1 minus the squared correlation function.
This implies that 
$g_Y\le1$ which is one among many other properties confirming that a DPP
is inhibitive, cf.\ \cite{LMR15} and \cite{biscio15}.

\subsection{Mat{\'e}rn hard core models of types I-II}\label{s:2.3}

Following \cite{matern:86} we define
 hard core point processes as follows.
Let $\Phi$ be a stationary Poisson process on $\mathbb R^d$ with
intensity $\rho_\Phi>0$, $D>0$ a hard core parameter, 
and $V=\{V(x):x\in\mathbb R^d\}$ a
 random process of independent
uniformly distributed random variables between 0 and 1, where
$\Phi$ and $V$
are independent. Denote 
$\omega_d=\pi^{d/2}/\Gamma(1+d/2)$ 
the volume of the $d$-dimensional unit ball, 
and $k_d(r,D)$ the volume of the intersection
of two $d$-dimensional
balls of radii $D$ and distance $r$ between their
centres.
The Mat{\'e}rn hard core model of type I, denoted $Y_I$, 
is given by the points in
$\Phi$ which are not $D$-close to some other point in $\Phi$, i.e.\ 
\[Y_I=\{x\in\Phi:\|x-y\|>D\mbox{ whenever $y\in\Phi\setminus\{x\}$}\}.\]
For the Mat{\'e}rn hard core model of type II, denoted $Y_{II}$, we
interpret $V(x)$ as the birth time
of $x$ and let $Y_{II}$ consist of the points 
$x\in\Phi$ such that no other $D$-close point in $\Phi$ is older than
$x$, i.e.\
\[Y_{II}=\{x\in\Phi:\|x-y\|>D\mbox{ whenever
  $y\in\Phi$ and $V(x)>V(y)$}\}.\]

These hard core point processes are stationary and isotropic with intensities 
\begin{equation}\label{e:rhoI-II}
\rho_{Y_I}=\rho_\Phi\exp\left(-\rho_\Phi\omega_dD^d\right), \quad
\rho_{Y_{II}}=\frac{1-\exp\left(-\rho_\Phi\omega_dD^d\right)}{\omega_dD^d},
\end{equation}
and pair correlation functions 
\begin{equation}\label{e:gI}
g_{Y_I,0}(r)=1(r>D)\exp(\rho_\Phi k_d(r,D))
\end{equation}
and
\begin{align}
g_{Y_{II},0}(r)=&1[r>D]
\frac{2\omega_dD^d}
{\left(\omega_dD^d-k_d(r,D)\right)
\left(1-\exp\left(-\rho_\Phi\omega_d D^d\right)\right)}\nonumber\\
&\left[1-
\frac{1-\exp\left(-\rho_\Phi\left(2\omega_dD^d-k_d(r,D)\right)\right)}
{\left(2\omega_dD^d-k_d(r,D)\right)
\left(1-\exp\left(-\rho_\Phi\omega_dD^d\right)\right)}\omega_dD^d
\right]\label{e:gII}
\end{align}
where $1(\cdot)$ is the indicator function. Note that
$Y_I\subseteq Y_{II}$ and
$\rho_{Y_I}<\rho_{Y_{II}}$. The pair correlation functions in
\eqref{e:gI}-\eqref{e:gII} are continuous except at $r=D$,
0 when $r\le D$, 
strictly decreasing for $r\in]D,2D[$, and 1 when $r\ge 2D$. 
 Finally, note that $k_d(r,D)=0$ if $r\ge 2D$, and 
\begin{equation}\label{e:kd}
k_d(r,D)=\left\{\begin{array}{lll}
 2D-r & \mbox{if $d=1$}\\
 2D^2\arccos\left(\frac{r}{2D}\right)-\frac{r}{2}\sqrt{4D^2-r^2} & \mbox{if $d=2$}\\
 \frac{4\pi}{3}D^3\left(1-\frac{3r}{4D}+\frac{r^3}{16D^3}\right) & \mbox{if $d=3$}
 \end{array}
\right.
\end{equation}
if $r\le 2D$.

\section{Specific models for the selection probabilities}\label{s:3}

Section~\ref{subs:general results} 
 discusses the implications of \eqref{e:rhonrhon} in general, 
while Sections~\ref{s:3.1}-\ref{s:3.2}  consider two classes of models for $\Pi$ where 
explicit expressions for our main 
characteristics ($q,\rho_X,M_0,g_{X,0}$) are available. 

\subsection{General results and conditions}\label{subs:general results}

In the remainder of this paper, to exclude non-interesting cases, we
focus on the following situation. 
We always assume that
$\Pi(o)$ has a positive variance or equivalently that $M_0(0)> 1$,
since we do not want $\Pi$ to be deterministic. 
In addition, we always assume that  
$\rho_X>0$ (or equivalently $\rho_Y>0$ and $q>0$), because otherwise   
 $X$ would be almost surely empty.
Since it is typically the case that an isotropic pair correlation
function tends to 1 as the distance tends to infinity, 
we want $M_0(r)$ to tend to 1 as $r\to\infty$, cf.\
\eqref{e:rhonrhon}. Therefore we are not so
interested in 
the case where $\Pi(x)$ does not depend on the location
$x$, since then $M_0$ is a
constant $\ge1$ and $\Pi$ is deterministic if $M_0=1$. 

We have $g_{X,0}=g_{Y,0}$ if and only if $\Pi$ is
uncorrelated, cf.\ \eqref{e:star} and \eqref{e:rhonrhon}. If $\Pi$ is
non-negatively correlated, i.e.\  $M_0\geq 1$, then $g_{X,0}\ge g_{Y,0}$, so $g_{Y,0}$
cannot cross 1 before 
$g_{X,0}$ crosses 1. 
If $\Pi$ is
positively correlated, i.e.\ $M_0>1$, then $g_{X,0}> g_{Y,0}$. If $\Pi$ can be
negatively correlated, a rather peculiar behaviour of
$g_{X,0}$ may happen and we shall exclude this case in our specific models.

By Cauchy-Schwartz inequality and since $\Pi^2\le\Pi$, we have for
$r=\|x_1-x_2\|\ge0$, 
\[M_0(r)=\frac{\mathrm E\left[\Pi(x_1)\Pi(x_2)\right]}{q^2}\le
\frac{\sqrt{\mathrm E\left[\Pi(x_1)^2\right]}
\sqrt{\mathrm E\left[\Pi(x_s)^2\right]}}{q^2}\le
\frac{\sqrt{\mathrm E\left[\Pi(x_1)\right]}
\sqrt{\mathrm E\left[\Pi(x_s)\right]}}{q^2}=\frac{1}{q}.\]
Combining this with \eqref{e:rhonrhon},
 we obtain an upper bound: $g_{X,0}\le g_{Y,0}/q$. 

Define $\tau_Y=\sup\{\tau\ge0: g_{Y,0}(r)=0\mbox{ whenever }r\le \tau\}$. 
When $\tau_Y>0$ we say that $Y$ is a hard-core process 
with hard-core parameter $\tau_Y$. 
Assume that $M_0(r)>0$
for all $r\ge0$; this is satisfied for all the models of
$\Pi$ specified later in this paper.
Then 
$\tau_X= \tau_Y$, cf.\ \eqref{e:rhonrhon}. Hence
$X$ is a hard-core process if and
only if $Y$ is a hard-core
process.

At the small scale, i.e.\ when $r\le \tau$ where $\tau>0$ is a 
sufficiently small constant, we have the
following.  
\begin{enumerate}
\item[(a)] Assume $M_0$ is continuous. Since $M_0(0)>1$, we can assume that
$M_0(r)>1$ for $r\le \tau$. Hence, by \eqref{e:rhonrhon}, for $r\le \tau$, either
$g_{X,0}(r)>g_{Y,0}(r)$ or $g_{X,0}(r)=g_{Y,0}(r)=0$. 
\item[(b)] Assume both $M_0$ and
$g_{Y,0}$ are continuous. Then $g_{X,0}$ is continuous, and so we can assume that
$g_{Y,0}\le g_{X,0}(r)<1$ for $r\le \tau$. Consequently, at distance
$r\le \tau$, the inhibitive behaviour 
of $Y$ 
(quantified in terms of its pair correlation function)
is preserved in $X$ but it cannot be stronger. 
\end{enumerate}
In brief we will 
be interested in models where $\Pi$ is positively and not too
weakly correlated at the small scale. 

At the large scale, basically the properties of $g_X$
depends on $q$ and the range of correlation of $\Pi$. 
If $q$ is large, then since 
$g_{X,0}\le g_{Y,0}/q$ we may have $g_X\le1$, 
meaning that
no clustering is created by the thinning process; an obvious
example is when $g_{Y,0}\le1$, as
e.g.\ in a DPP.
If $q$ is sufficiently small, then  
$\sup_r g_X(r)>1$ occurs in our examples of models, 
and we expect this to be the situation in many other cases. 
However,  it is not true that there always exists $q$ such that  
$\sup_r g_X(r)>1$. An obvious counterexample is when $\Pi$ is
uncorrelated; other counterexamples may be constructed 
when the variance $\sigma^2$, say, of $\Pi$ is 
such that $\sigma^2/q^2\to 0$ as $q\to 0$.
 On the other hand, assume $q$ is fixed and $\Pi$ is non-negatively
 correlated, then it is always possible to get  $\sup_r g_X(r)>1$ by
 increasing the range of correlation of $\Pi$, i.e.\ making $M_0(r)>1$
 for $r$ sufficiently large. This is exemplified in 
Sections~\ref{s:3.1}-\ref{s:3.2}.  

\subsection{Transformed Gaussian processes}\label{s:3.1}

This section assumes $-\log\Pi$ is the $\chi^2$-process given by
\begin{equation}\label{pi-gauss} 
\Pi(x)=\exp\left(-\frac 12\sum_{i=1}^k Z_i(x)^2\right),\quad
x\in\mathbb R^d,\end{equation} 
where $Z_i=\{Z_i(x):x\in\mathbb R^d\}$,
$i=1,\ldots,k$, are i.i.d.\ zero-mean real Gaussian processes
with covariance function $K:\mathbb R^d\times \mathbb R^d\mapsto 
\mathbb R$. 

A straightforward calculation yields $\mathrm E[\Pi(x)]=1/(1+K(x,x))^{k/2}$ and
\[M(x_1,x_2) 
=\left[\frac{(1+K(x_1,x_1))(1+K(x_2,x_2))}
{(1+K(x_1,x_1))(1+K(x_2,x_2))-|K(x_1,x_2)|^2}\right]^{k/2}.\]
Hence, for $K(x_1,x_1)K(x_2,x_2)>0$ and
 defining $R(x_1,x_2)=K(x_1,x_2)/\sqrt{K(x_1,x_1)K(x_2,x_2)}$, 
\[M(x_1,x_2) 
=\left[1-\frac{R(x_1,x_2)^{2}}{(1+1/K(x_1,x_1))(1+1/K(x_2,x_2))}\right]^{-k/2}\]
is an increasing function of
$|R(x_1,x_2)|,K(x_1,x_1),K(x_2,x_2)$, respectively. 

In the sequel we
assume stationarity and isotropy of $K(x_1,x_2)=K_0(\|x_1-x_2\|)$, whereby 
$\Pi$ is stationary and isotropic. 
Defining
$\kappa=K_0(o)$ and $R_0=K_0/\kappa$, 
we notice as the variance $\kappa$ increases from zero to infinity that
\begin{equation}\label{eq:qkappa}q=1/(1+\kappa)^{k/2}\end{equation}
decreases from 1 to 0, while for fixed $r$,
\begin{equation}\label{e:a}
M_0(r) 
=\left[1-\frac{R_0(r)^2}{(1+1/\kappa)^2}\right]^{-k/2} = \left[1-(1-q^{2/k})^2 R_0(r)^2\right]^{-k/2} 
\end{equation}
increases from 1 to $[1-R_0(r)^2]^{-k/2}$. Thus there is a trade-off
between how large $q=\rho_X/\rho_Y$ and $M_0(r)=g_{X,0}(r)/g_{Y,0}(r)$ can be.

We have that $M_0(r)\geq 1$ is a decreasing function of $k$ and $M_0(r)\to 1$ as $k\to\infty$, showing that taking a large value of $k$ is not appropriate if we want  $X$ to exhibit a clustering behaviour at the large scale. 
Further, assume that the correlation function $R_0$ depends on a scale
parameter $s>0$, i.e.\ for all $r\geq 0$, $R_0(r)=R_0(r,s)=\tilde
R_0(r/s)$ where $\tilde R_0(r)=R_0(r,1)$. This is so for most
parametric models of covariance functions used in spatial statistics. Then, for any given $q\in (0,1)$ and $k\geq 1$, we have $M_0(r)\to [1-(1-q^{2/k})^2]^{-k/2}>1$ as $s\to\infty$ provided $\tilde R_0$ is continuous at the origin. This combined with \eqref{e:rhonrhon} proves that $X$ will necessarily exhibit some clustering behaviour at the large scale when $s$ is sufficiently large.

\begin{figure}[h]
\begin{center}
\begin{tabular}{ccc} 
\includegraphics[scale=0.18]{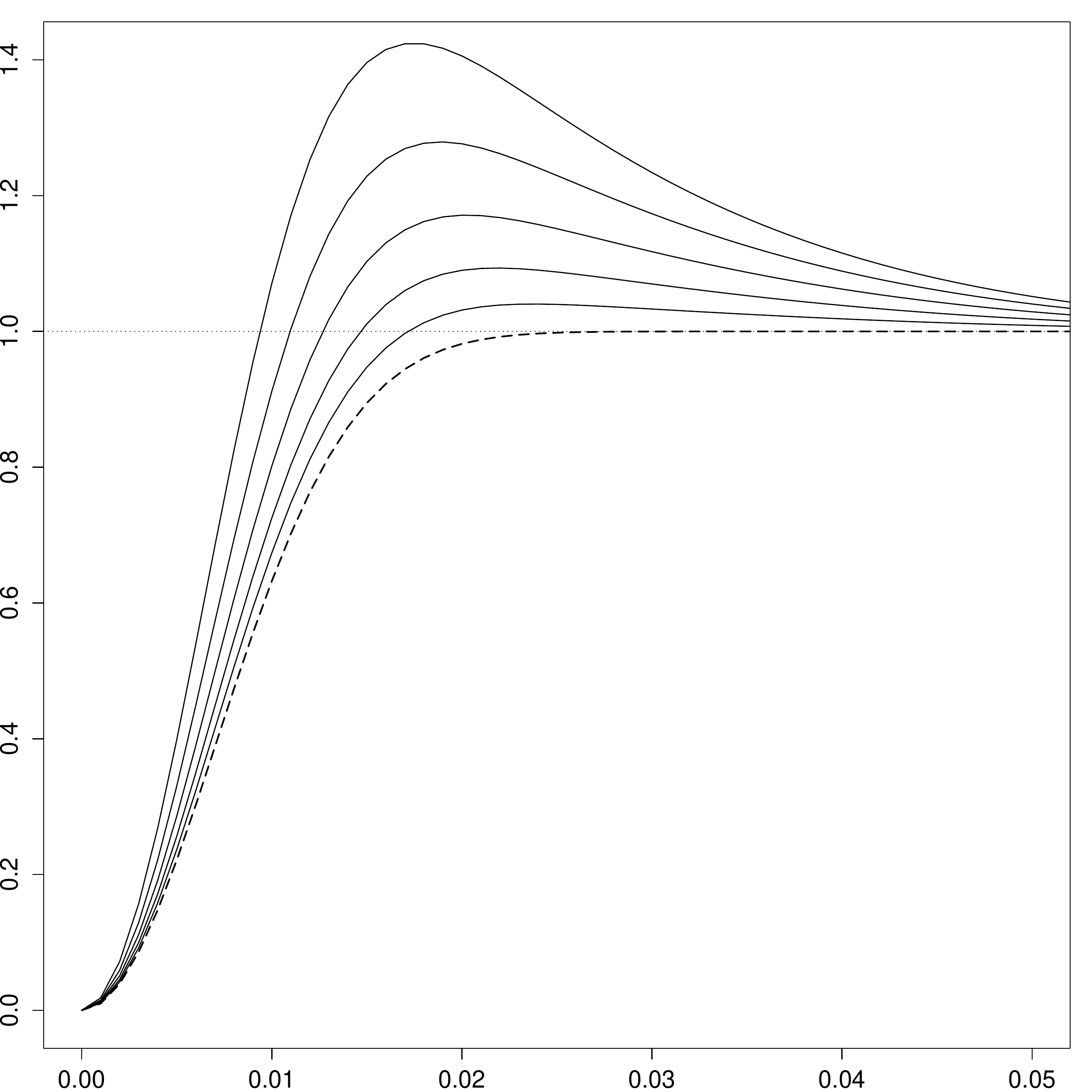}  & \includegraphics[scale=0.18]{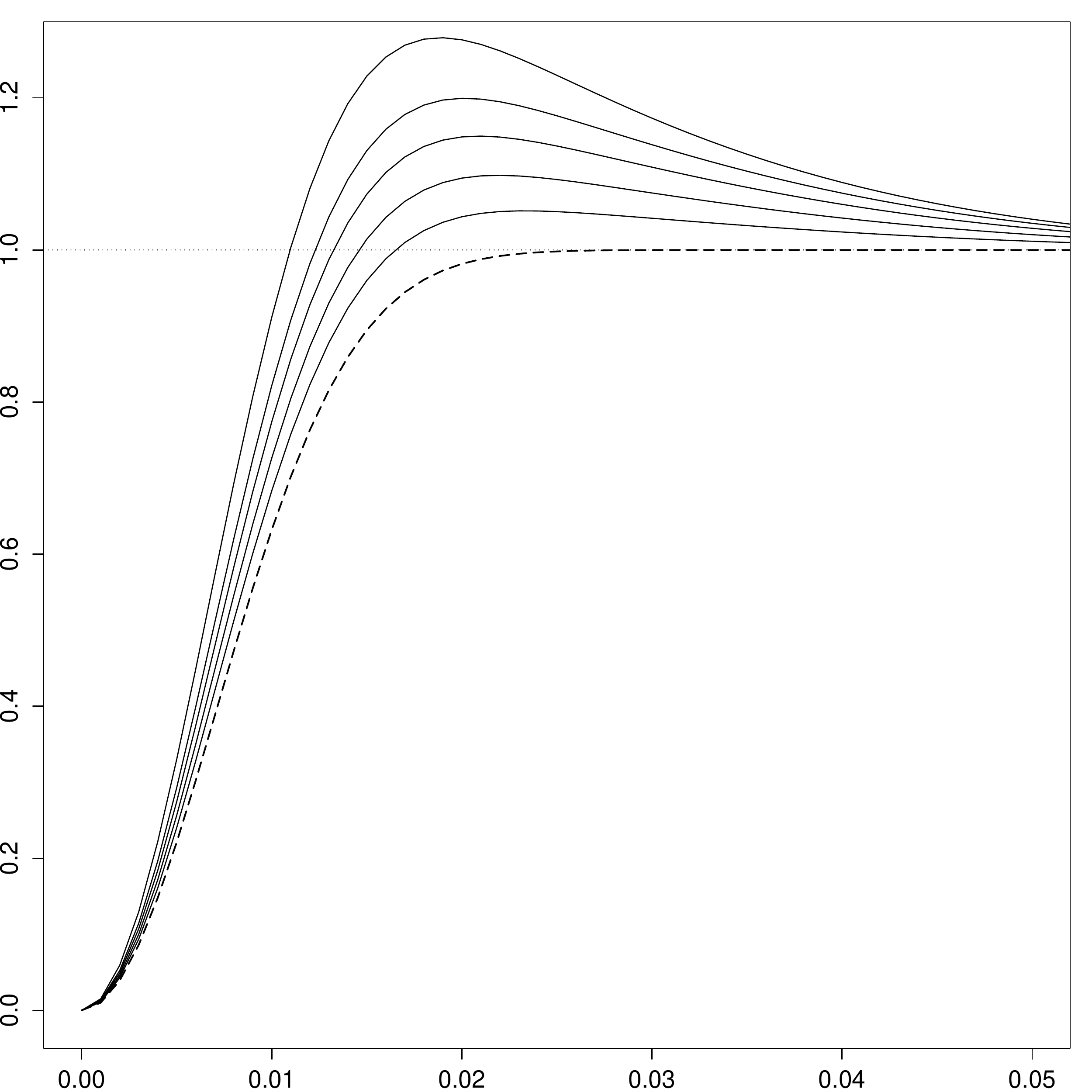} & \includegraphics[scale=0.18]{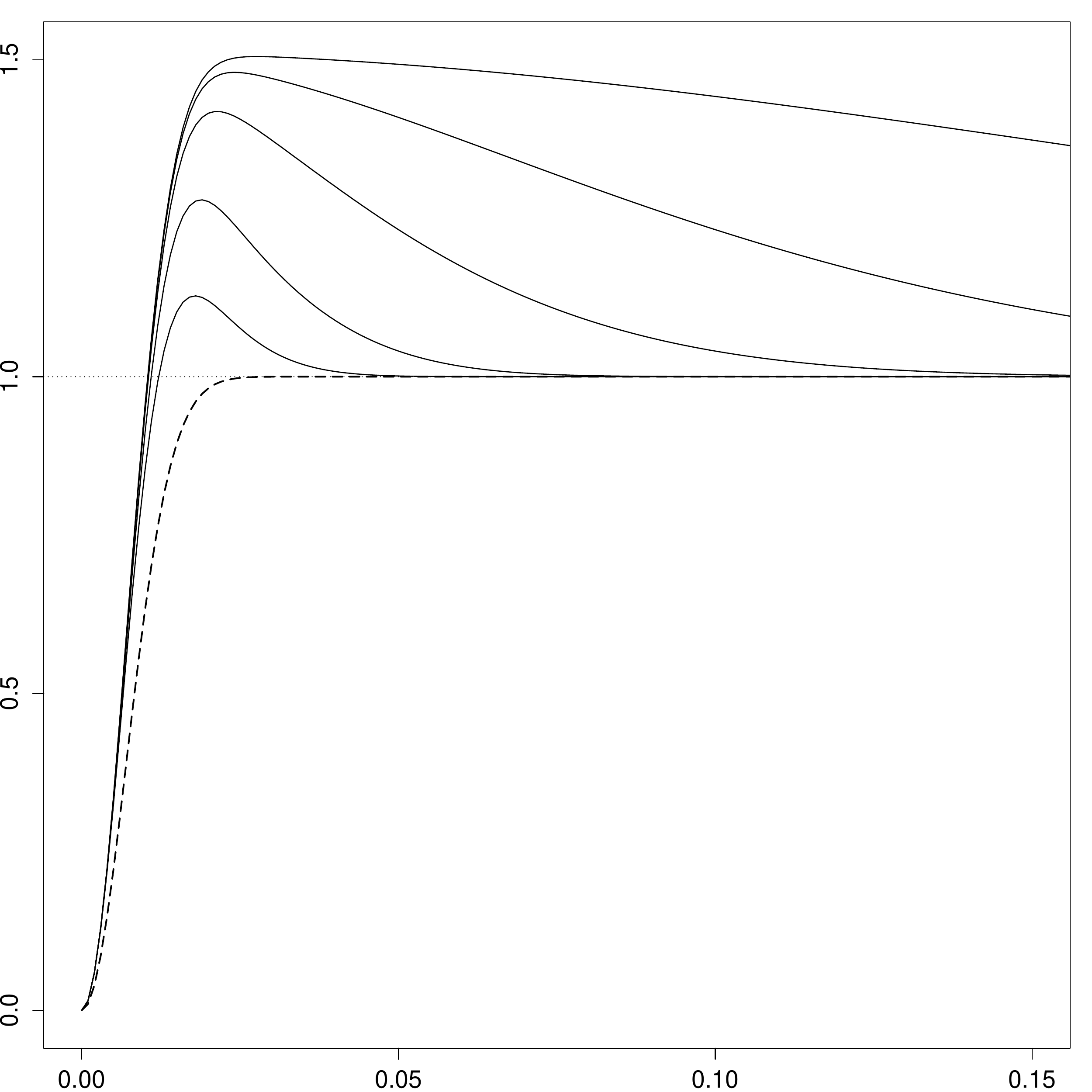}\\
 \includegraphics[scale=0.18]{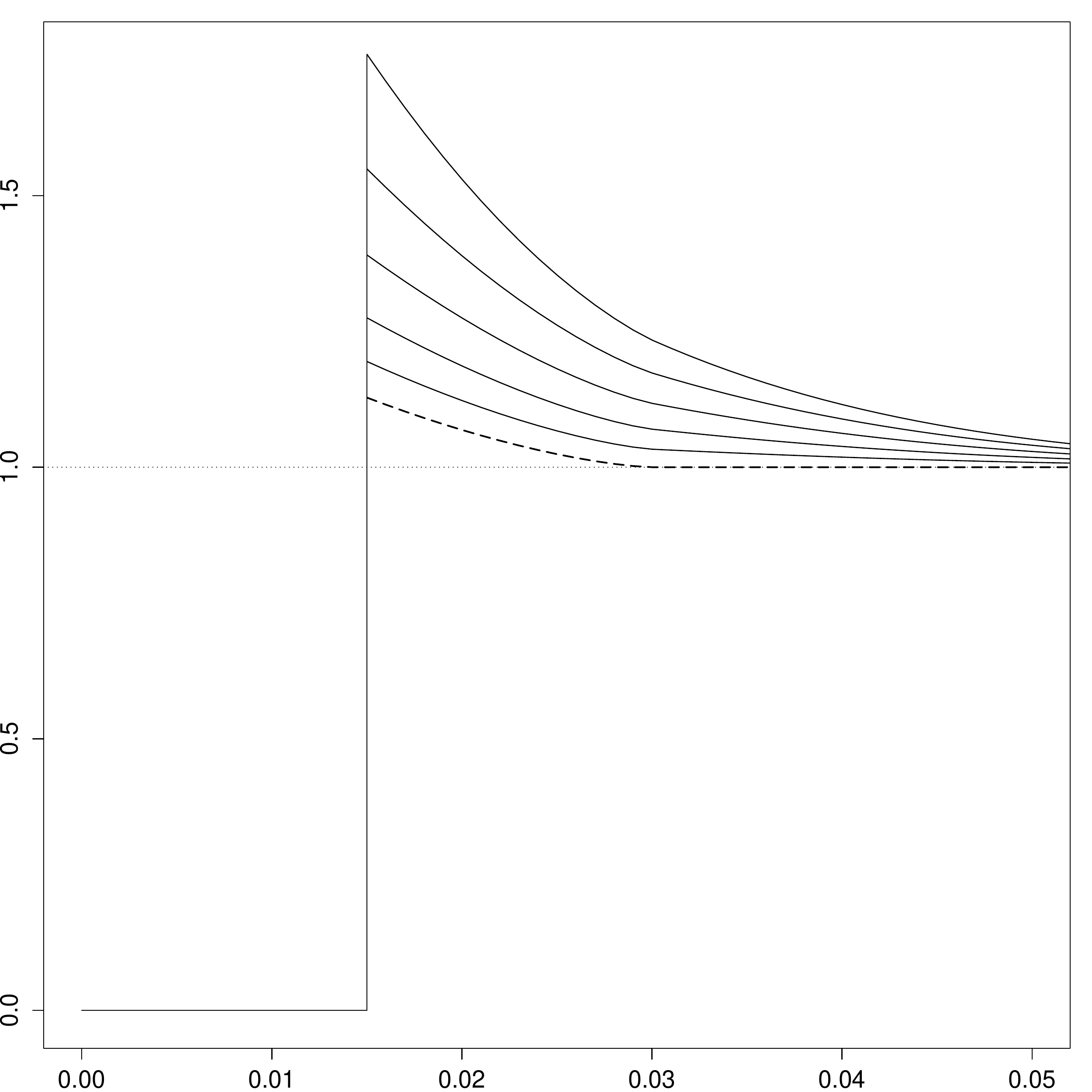}  & \includegraphics[scale=0.18]{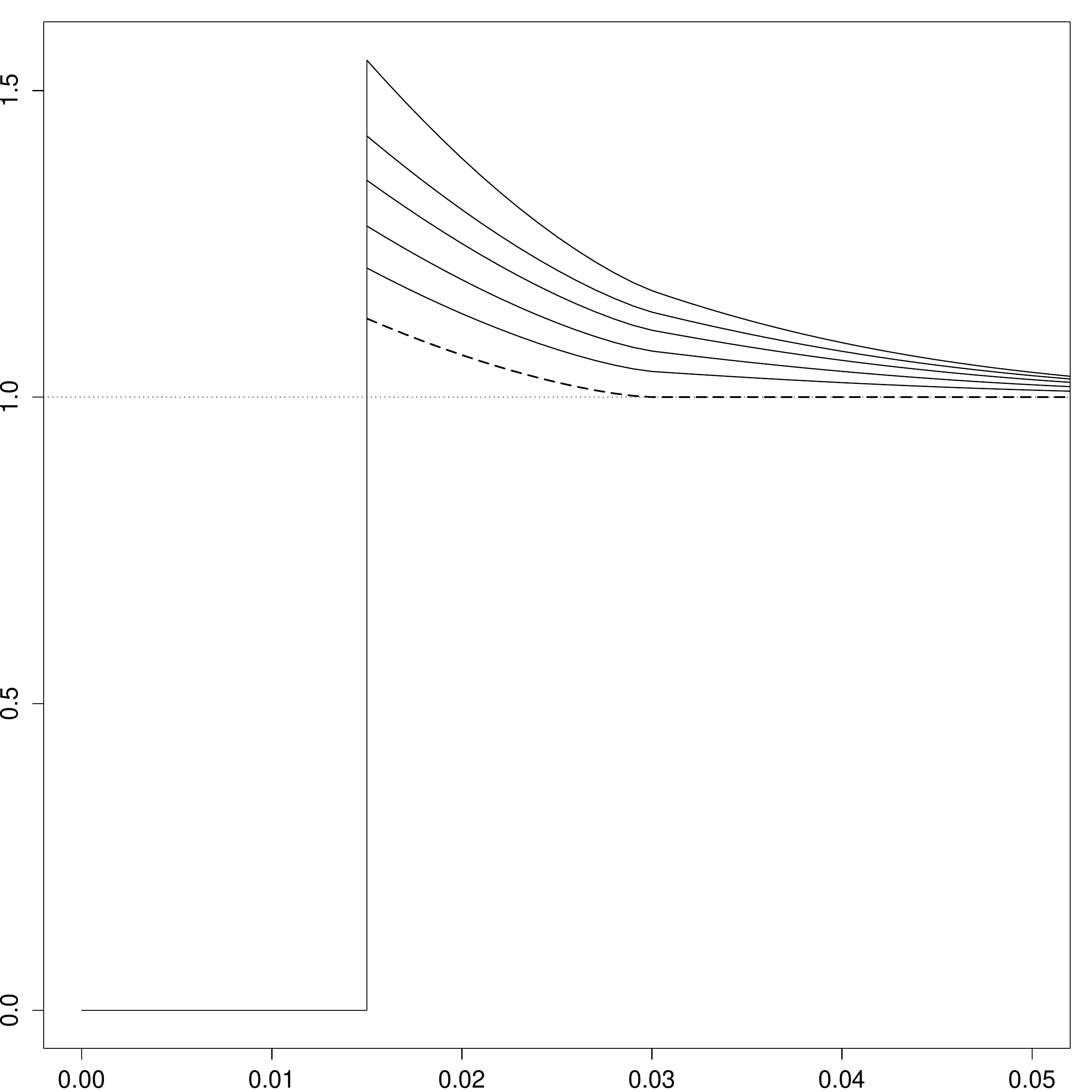} & \includegraphics[scale=0.18]{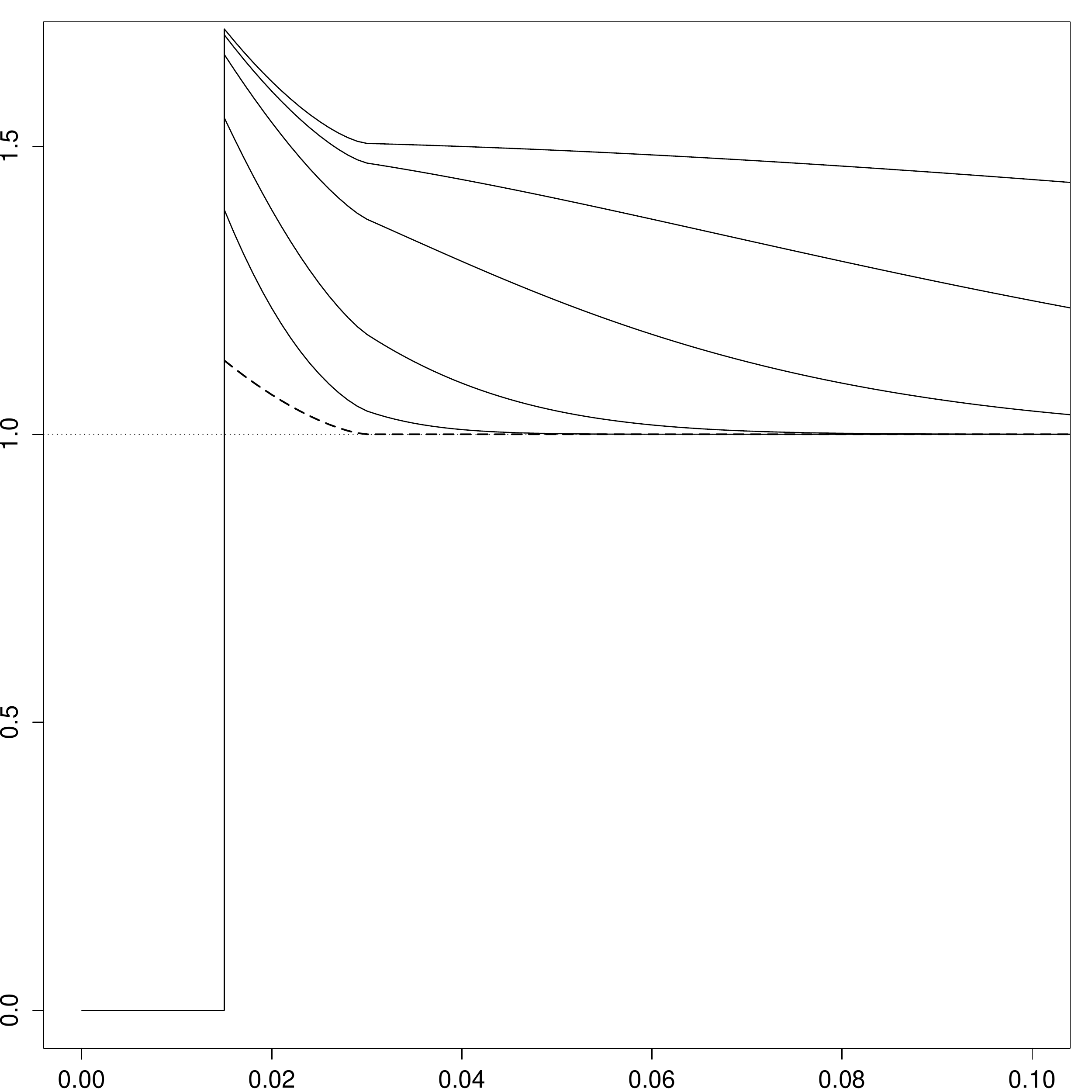}\\
\end{tabular}
\caption{Pair correlation functions of $Y$ (dashed line) and $X$
  (solid lines) when $Y$ is a DPP with a Gaussian kernel (first row)
  or a Mat{\'e}rn hard core process of type II (second row) 
and $-\log\Pi$ is the $\chi^2$-process given by~\eqref{pi-gauss} 
when $K_0$ is a Gaussian covariance function with scale parameter
$s$. First column: $k=1$, $s=0.05$, and from top to bottom $q=0.4,0.5,0.6,0.7,0.8$; second column: $q=0.5$, $s=0.05$, and from top to bottom $k=1,2,3,5,10$; third column:  $q=0.5$, $k=1$, and from top to bottom $s=0.5,0.2,0.1,0.05,0.03$.}\label{fig:pcf-gauss}
\end{center}
\end{figure}

\begin{figure}[h]
\begin{center}
\begin{tabular}{cc} 
\includegraphics[scale=0.3]{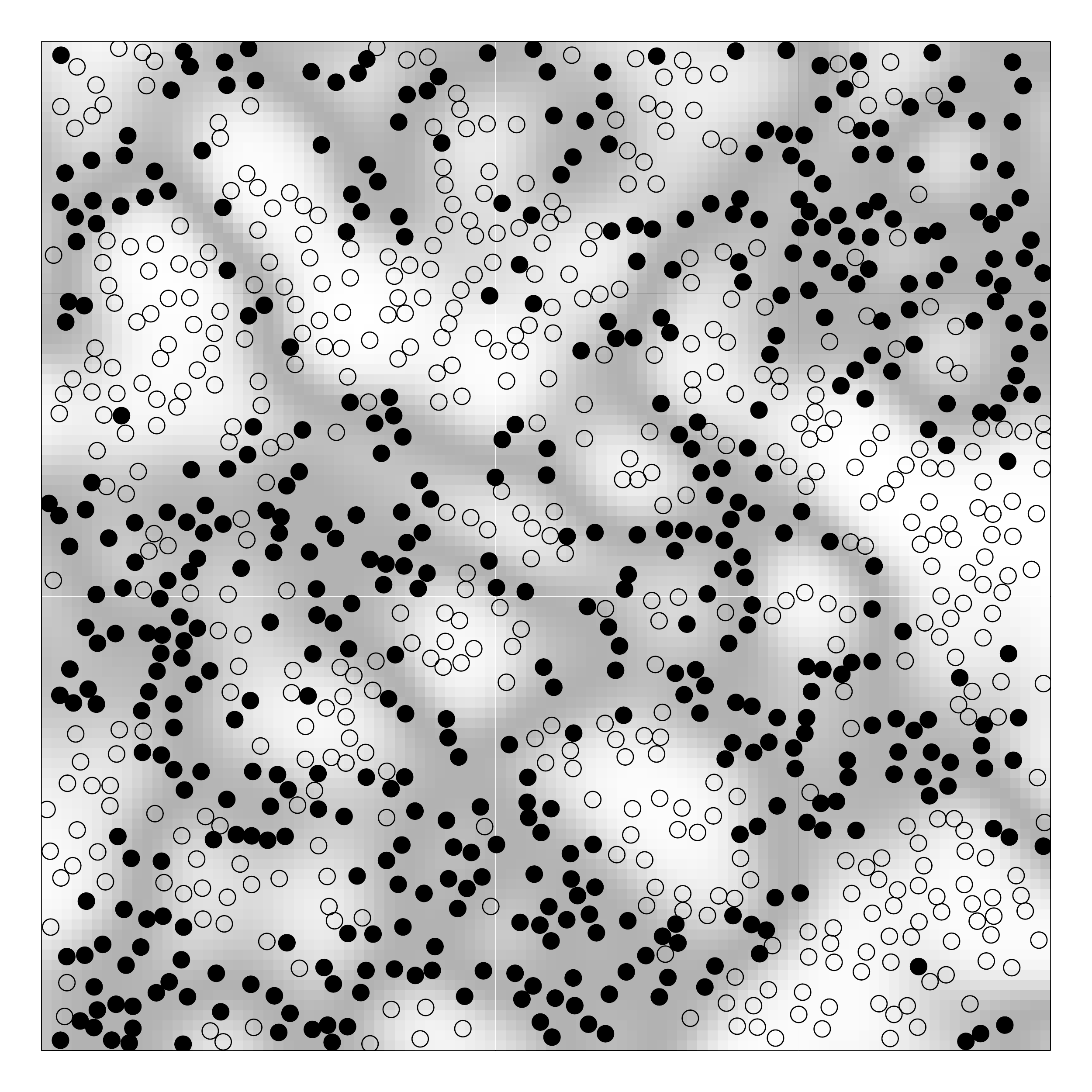}  & \includegraphics[scale=0.3]{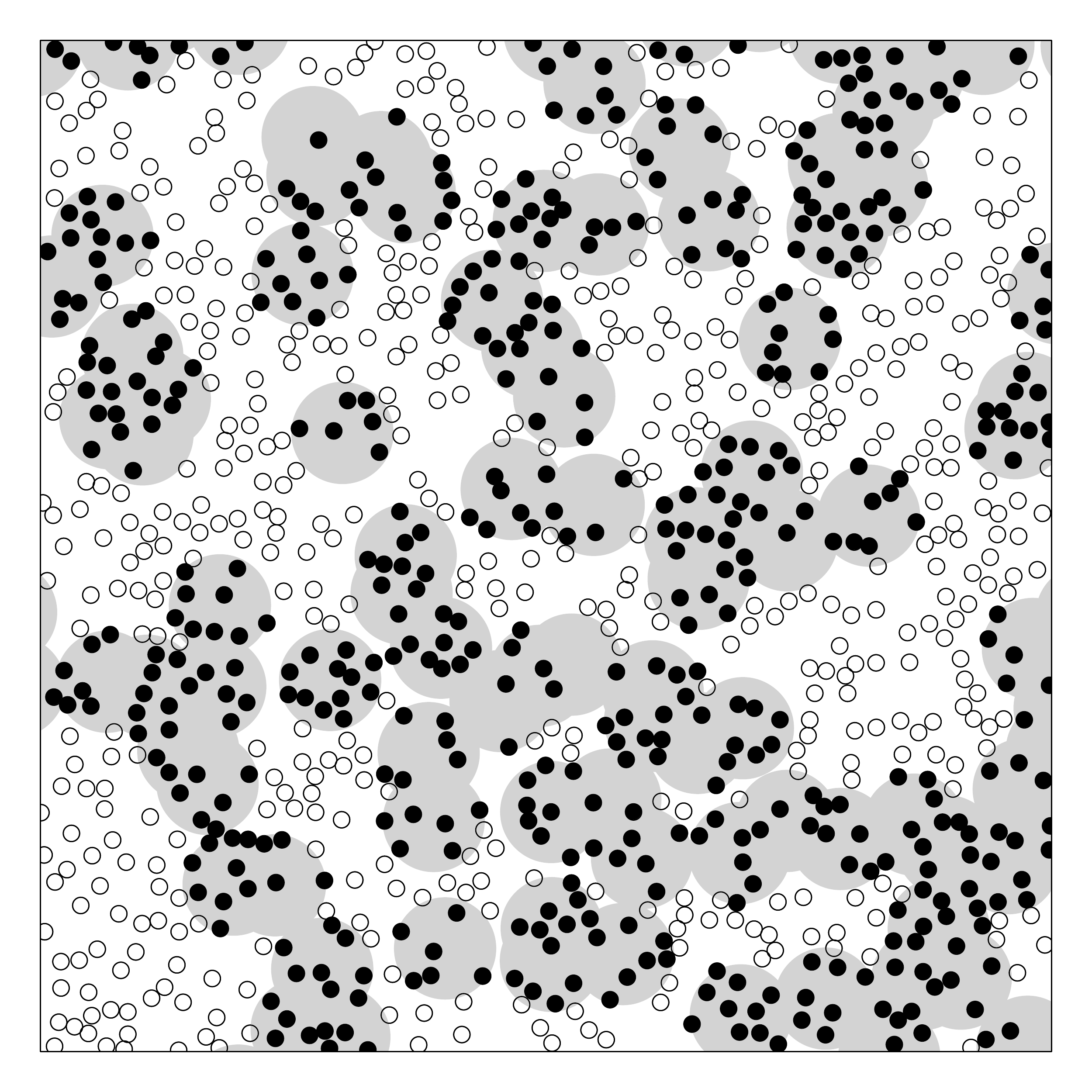} \\
\end{tabular}
\caption{Realization of $Y$ (union of all points) and $X$ (black points)
  within a unit square when $Y$ is a Mat{\'e}rn hard core
  process of type II 
 with $\rho_Y=1000$ and $D=0.015$. Left: Corresponding realization of
  $\Pi$ (the grayscale from white to black corresponds to 
  values from 0 to 1) when $-\log\Pi$ is the
$\chi^2$-process given by~\eqref{pi-gauss} when $k=1$ and $K_0$ is
  the Gaussian covariance function with scale parameter
  $s=0.1$. Right: Corresponding realization of
  $\Pi$ represented by the characteristic function for
  the union of gray discs within the unit
  square and specified by the Boolean disc model with fixed radius
  $\Delta_0=0.05$, cf.~\eqref{pi_bool}. For both plots  $q=0.5$.}\label{fig:simus_XYPI}
\end{center}
\end{figure}

The effect of the parameters is illustrated in
Figure~\ref{fig:pcf-gauss} which shows the pair correlation of $X$
when $d=2$ and $Y$ is either a DPP or a type II Mat{\'e}rn hard core
process. Specifically, the first row of Figure~\ref{fig:pcf-gauss}
corresponds to the case where $Y$ is a DPP with a Gaussian kernel
$C_0(r)=\rho_Y \exp(-(r/0.015)^2)$ and $\rho_Y=1000$, while in the
second row $Y$ is a type II Mat{\'e}rn hard core process with
$D=0.015$ and $\rho_\Phi=1736$ whereby $\rho_Y=1000$. The selection
probabilities are given by~\eqref{pi-gauss} where $K_0$ is a Gaussian
covariance function with scale parameter $s$. A joint realization of
the restrictions of $Y$, $\Pi$, and $X$ to a unit square 
is shown on the left hand side of Figure~\ref{fig:simus_XYPI}.

\subsection{Boolean and complementary Boolean models}\label{s:3.2}

This section specifies further models 
for the selection probabilities.

Let $\Psi$ be a stationary Poisson process on $\mathbb R^d$ with
intensity $\rho_\Psi$, and conditional on $\Psi$, let $\Delta_0$ and $\Delta_x$
for all $x\in\Psi$ be i.i.d.\ positive
random variables with a distribution which does not depend on
$\Psi$ and so that $\mathrm E[\Delta_0^d]<\infty$. 
Denote $\Xi$ the stationary Boolean model given by the union of the $d$-dimensional
 balls centred at the points of
$\Psi$ and with radii $\Delta_x$, 
$x\in\Psi$. 
Recall that 
$p=P(o\in \Xi)$ is the volume fraction, and for $x\in\mathbb R^d$ and $r=\|x\|$,
$C_\Xi(r)=P(\{o,x\}\subset \Xi)$ is the so-called covariance
  function, where expressions for $p,C_\Xi(r)$, and the void probability 
$P(\{o,x\}\cap \Xi=\emptyset)$ are known (see e.g.\ \cite{molchanov:97}).

Specifying $\Pi$ by the characteristic function of the random set $\Xi$ 
or its complement $\Xi^c$, in either case $X$ becomes
stationary and isotropic: First, if 
 \begin{equation}\label{pi_bool}\Pi(x)=1(x\in \Xi),\end{equation}  
then
\begin{equation}\label{e:h}
q=p=1-\exp\left(-\rho_\Psi\omega_d\mathrm E\left[\Delta_0^d\right]\right)
\end{equation}
and since $\mathrm E\left[\Pi(o)\Pi(x)\right]=C_\Xi(\|x\|)$, we obtain
\begin{equation}\label{e:j}
M_0(r)=\frac{2}{p}-\frac{1}{p^2}+\left(\frac{1-p}{p}\right)^2
\exp\left(\rho_\Psi\mathrm E\left[k_d(r,\Delta_0)\right]\right).
\end{equation}
Second, if $\Pi(x)=1(x\not\in \Xi)$, then by \eqref{e:rhonrhon_comp},
\begin{equation}\label{e:k}
q=1-p=\exp\left(-\rho_\Psi\omega_d\mathrm E\left[\Delta_0^d\right]\right)
\end{equation}
and 
\begin{equation}\label{e:l}
M_0(r)=\exp\left(\rho_\Psi\mathrm E\left[k_d(r,\Delta_0)\right]\right).
\end{equation}
Equations~\eqref{e:h}-\eqref{e:l} become explicit in the particular
case of a fixed deterministic radius $\Delta_0>0$.
When $\Delta_0$ is random, $\mathrm
E\left[k_d(r,\Delta_0)\right]$ may be evaluated by a numeric method
using \eqref{e:kd}.
We consider later the case where $\Delta_0$ follows a Beta-distribution with
parameters $\alpha$ and $\beta$; then $\mathrm
E\left[\Delta_0^d\right]=B(\alpha+d,\beta)/B(\alpha,\beta)$ is given
in terms of the beta-function.

\begin{figure}[h]
\begin{center}
\begin{tabular}{cc} 
\includegraphics[scale=0.22]{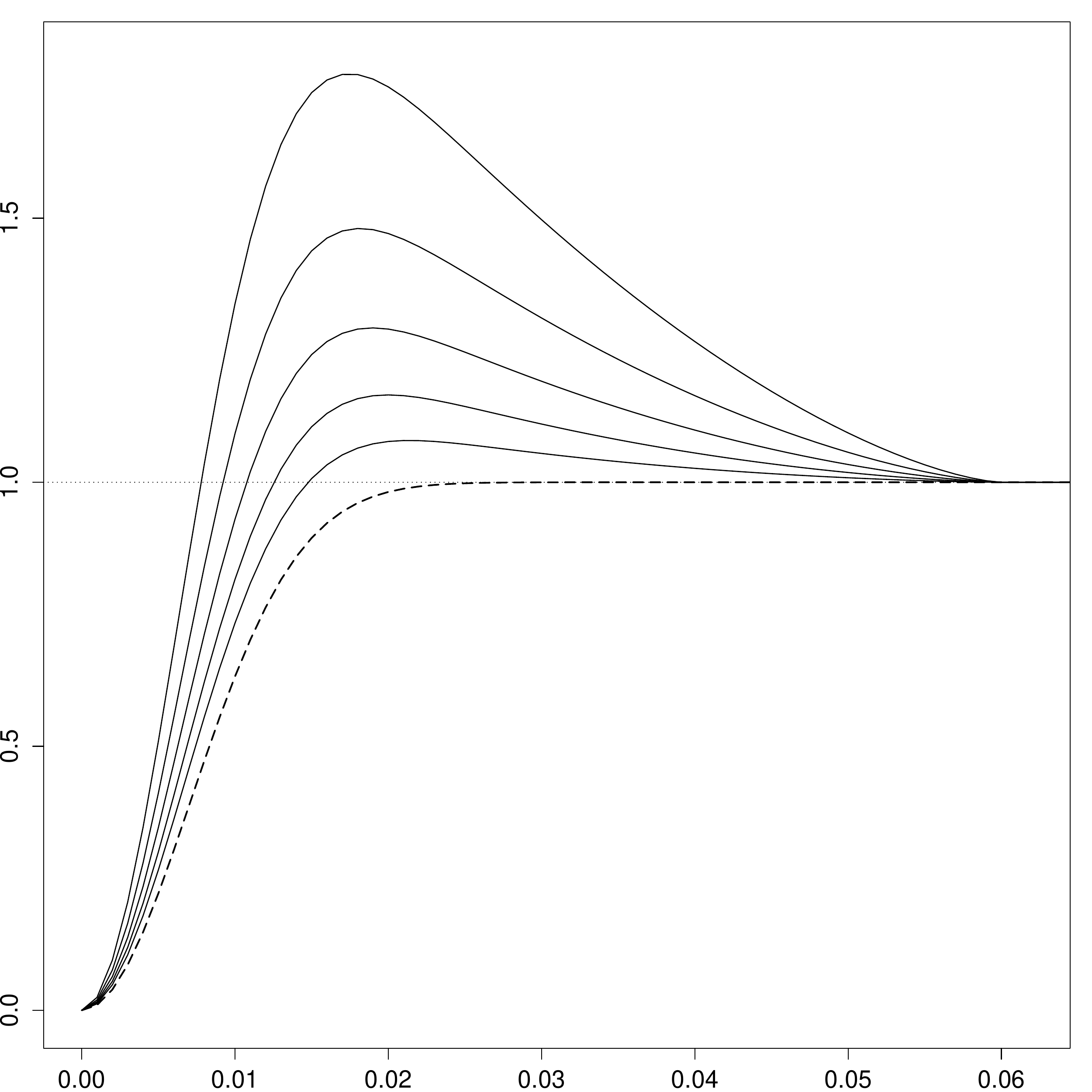}  & \includegraphics[scale=0.22]{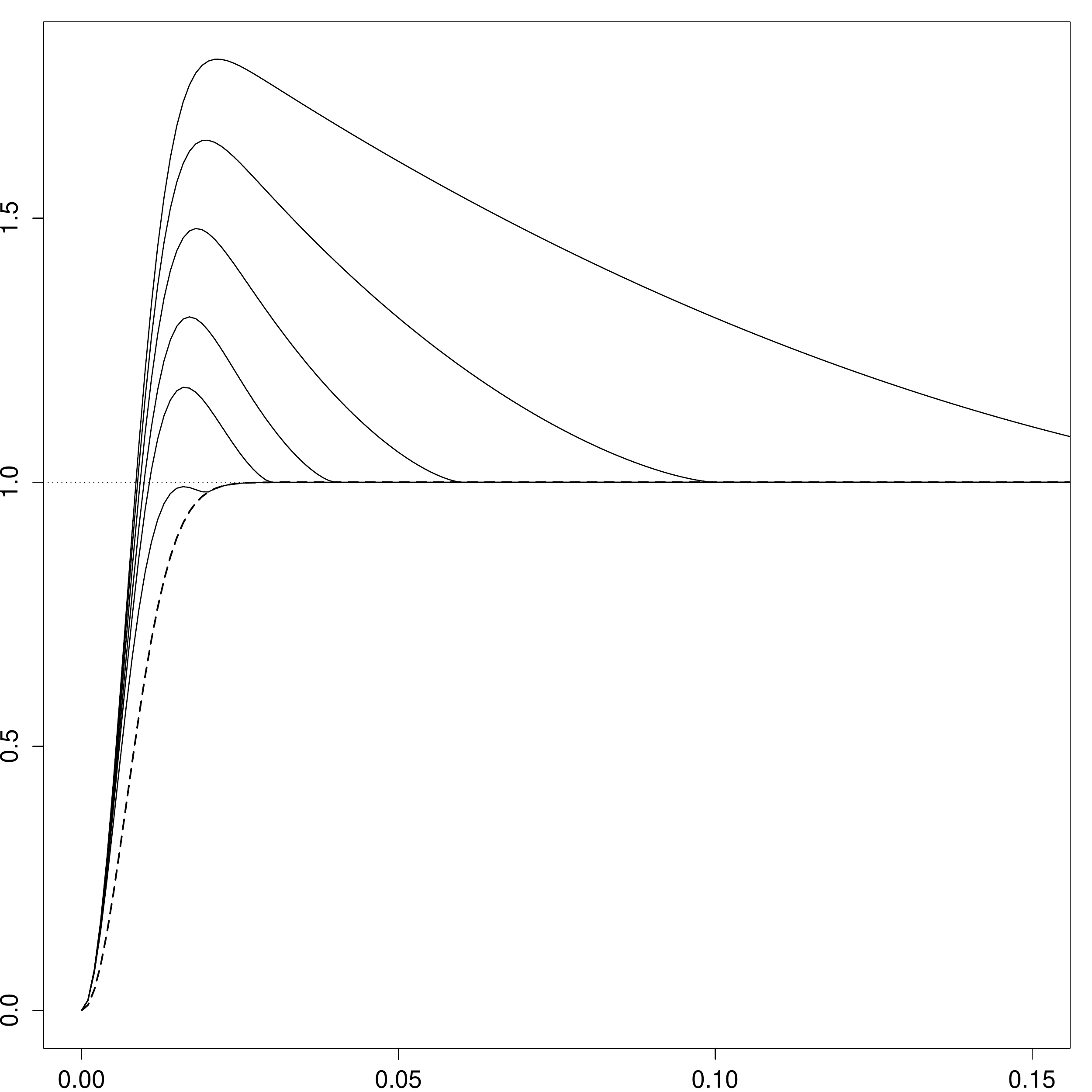} \\
 \includegraphics[scale=0.22]{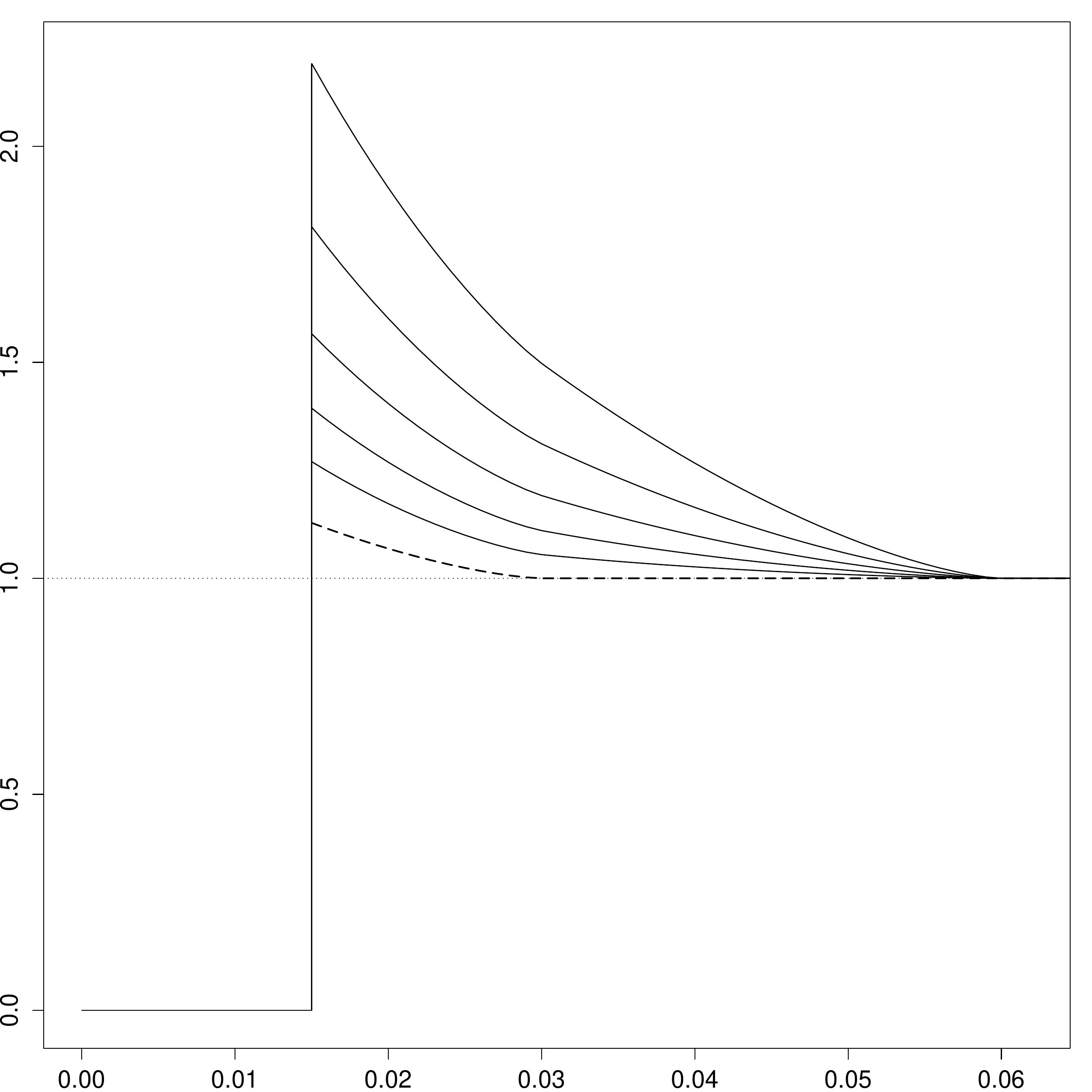}  & \includegraphics[scale=0.22]{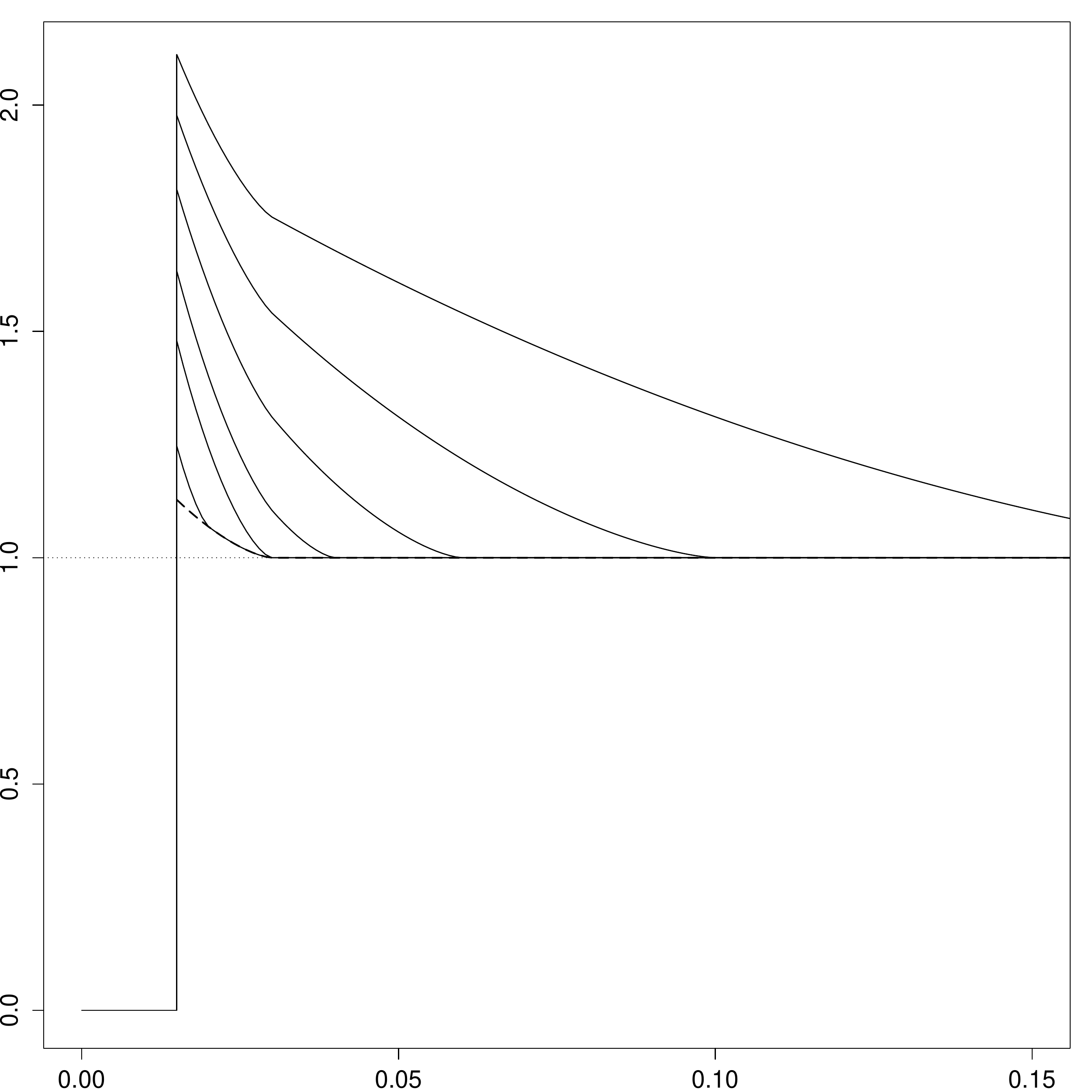} \\
\end{tabular}
\caption{Pair correlation functions of $Y$ (dashed line) and $X$
  (solid line) when $Y$ is a DPP with a Gaussian kernel (first row) or
  a Mat{\'e}rn hard core process of type II (second row) and $\Pi$ is the Boolean inclusion probability model given by~\eqref{pi_bool} with a deterministic radius $\Delta_0$. First column:  $\Delta_0=0.03$ and from top to bottom $q=0.4,0.5,0.6,0.7,0.8$; Second column:  $q=0.5$ and from top to bottom $\Delta_0=0.1,0.05,0.03,0.02,0.015,0.01$.}\label{fig:pcf-bool}
\end{center}
\end{figure}

Note that $M_0(r)\to 1/q>1$ as $\mathrm E\left[\Delta_0^d\right]\to
\infty$, showing that $X$ will necessarily exhibit some clustering
behaviour at the large scale if the Boolean model has large radii. The
pair correlation function of $X$ is represented in
Figure~\ref{fig:pcf-bool} for different values of the parameters in
the situation where $Y$ is either a Gaussian DPP or a type II
Mat{\'e}rn hard core process as in Figure~\ref{fig:pcf-gauss}, and
$\Pi$ is given by~\eqref{pi_bool} with a deterministic radius
$\Delta_0$. A joint realization of the restrictions of 
$Y$, $\Pi$ and $X$ to a unit square is shown on the right-hand side of Figure~\ref{fig:simus_XYPI}.

Finally, we notice that 
another tractable model for $\Pi$ is the characteristic function for a
random closed set given by the excursion set for a Gaussian process,
where a relation between $M_0$ and the covariance function of the
Gaussian process can be established, see 
\cite{chiu:stoyan:kendall:mecke:13} and the references
therein. 

\section{Simulation and inference}\label{s:inference}

In the sequel $W\subset\mathbb R^d$ denotes a bounded region (e.g.\ an
observation window). 
Section~\ref{s:simulation} concerns simulation of $(Y,\Pi,X)$ on
$W$ and conditional
simulation of $\Pi$ given a realization of $X$ on $W$.
Section~\ref{s:methods} deals with parametric inference methods
depending on whether we observe both $X$ and $Y$ on $W$  
or only $X$ on $W$, and Section~\ref{s:simu} discusses a simulation
study for these two cases.

\subsection{Simulation and conditional simulation}\label{s:simulation}

Simulating  $X$ within $W$ is straightforward from its
definition \eqref{e:th} as long as we are able to simulate the
restrictions of $Y$ and
$\Pi$ to $W$. Concerning our examples of $Y$, an algorithm to
generate a DPP within a rectangular window is detailed in \cite{LMR15}
while a Mat{\'e}rn hard core process of type I or II 
is easily simulated within any
bounded region.  For both models, some simulation routines are
available in the \texttt{spatstat} library \citep{baddeley:turner:05}
of  R \citep{R:14}.  Concerning $\Pi$, 
simulating the model in Section~\ref{s:3.1} amounts to simulate  a
centered Gaussian process with prescribed covariance function, which is
for instance implemented in the \texttt{RandomFields} library
\citep{RandomFields}, while 
generating a Boolean disc model for the example of Section~\ref{s:3.2} is straightforward. 

Suppose we have fitted a model for $\Pi$, based on the
observation of $X$ on $W$ (e.g.\ using the method
described in Section~\ref{s:methods}), and we are  interested in the
conditional simulation of $\Pi$ (possibly restricted to $W$)
given the observed point pattern $X_W=x_W$. 
This amounts to simulate according to the distribution of $\Pi$ given
$X_W=x_W$. 
The conditional distribution of $X_W$ given $\Pi$ and $Y_W$ 
admits the probability mass function 
\begin{equation}\label{density}
p(x_W | \Pi,Y_W) = 1(x_W\subseteq Y_W)\left\{\prod_{u\in x_W} \Pi(u)\right\}\left\{\prod_{v\in Y_W\setminus x_W} (1-\Pi(v))\right\},
\end{equation}
so the conditional distribution of $\Pi$ given $X_W=x_W$ is 
\begin{equation}\label{Pi_given_X}
 \mathrm P(\Pi\in F|X_W=x_W) \propto
\mathrm E\left[1(\Pi\in F,\,x_W\subseteq Y_W) \left\{\prod_{u\in x_W} \Pi(u)\right\}\left\{\prod_{v\in Y_W\setminus x_W} (1-\Pi(v)) \right\}\right],
 \end{equation}
where the constant of proportionality depends only on $x_W$.
The conditional simulation of $\Pi$ given $X_W=x_W$ would  thus
require some Monte Carlo based algorithm such as the
Metropolis-Hastings algorithm in order to approximate the expectation in \eqref{Pi_given_X}.  This is in general prohibitively time consuming and we do not consider this conditional simulation in the following. 

A simpler setting occurs when both $X$ and $Y$ are observed on
$W$. Let $\bar X_W = Y_W \setminus X_W$. Since $Y$ and $\Pi$ are independent,
the conditional distribution of $\Pi$ given $X_W=x_W$ and $\bar X_W=\bar x_W$ is
\begin{equation}\label{Pi_given_XY}
 \mathrm P(\Pi\in F|X_W=x_W,\bar X_W=\bar x_W) \propto 
\mathrm E\left[1(\Pi\in F)\left\{\prod_{u\in x_W} \Pi(u)\right\} 
\left\{\prod_{v\in \bar x_W} (1-\Pi(v))\right\}
\right].
 \end{equation}
The expectation in \eqref{Pi_given_XY} is simpler than that in
\eqref{Pi_given_X} but in general some 
Monte Carlo based algorithm is still needed for conditional
simulation. 
We detail two convenient situations below. 

The first case occurs when $\Pi$ is given by the Boolean model \eqref{pi_bool}. Then simulating according to  \eqref{Pi_given_XY} just reduces to the
conditional simulation of a Boolean random set $\Xi$ given that
$x_W\subseteq \Xi$ and $\bar x_W\cap \Xi=\emptyset$. This case of
conditional simulation is well known, see \cite{lantuejoul2002}.

In the second case,  $-\log\Pi$ is the
square of a stationary and isotropic Gaussian process given by
\eqref{pi-gauss} with $k=1$.  Then conditional simulation of $\Pi$
based on \eqref{Pi_given_XY}  amounts to generate the Gaussian process $Z$ given $X_W=x_W$ and $\bar X_W=\bar x_W$, which can be 
conducted  in two steps. 
In the first step, as described below 
generate $(Z(y_1),\dots,Z(y_n))$ given that
$X_W=x_W$, $\bar X_W=\bar x_W$, and $Y_W=x_W\cup\bar
x_W=\{y_1,\dots,y_n\}$, say. In the second step, 
simulate $Z$ on $W$, conditional on the values of $Z(y_i)$,
$i=1,\dots,n$,  generated in  the first step. This second step can be
done by double kriging as explained in \cite{lantuejoul2002} and this
is implemented in the \texttt{RandomFields} library of R. For the
first step, denote the number of points in $x_W$ by $n_x=n(x_W)$, and
similarly let $n_{\bar x}=n(\bar x_W)$ so that $n=n_x+n_{\bar x}$, and let 
$\Gamma$ be the $n\times n$
matrix with generic element $K_0(\|y_i-y_j\|)$. Assuming $\Gamma$ is invertible,
we deduce from  \eqref{Pi_given_XY} that the target law admits a density in $\mathbb  R^n$ of the form $f(\mathbf z)=c \exp(-U(\mathbf z))$, where $c>0$,
$\mathbf z=(z_1,\dots,z_n)$, and 
\[U(\mathbf z)=\frac 12 \sum_{i=1}^{n_x} z_i^2 - \sum_{i=n_x+1}^{n} \log(1-{\rm e}^{-\frac 12 z_i^2}) + \frac 12 \mathbf z \Gamma^{-1} \mathbf z',\]
where $\mathbf z'$ is the transpose of $\mathbf z$. The conditional simulation of
$(Z(y_1),\dots,Z(y_n))$ can then be done by a Metropolis within Gibbs
sampler as follows, where $\mathcal N(0,\kappa)$ denotes the centered normal
distribution with variance $\kappa=K_0(0)$:  
\begin{enumerate}
\item generate $\mathbf z=(z_1,\ldots,z_n)$ as $n$ independent $\mathcal
  N(0,\kappa)$-distributed random variables;
\item for $i=1$ to $n$

\hspace{0.5cm} let $\tilde z_i\sim \mathcal N(0,\kappa)$, $\tilde {\mathbf z}=(z_1,\dots,z_{i-1},\tilde z_i,z_{i+1},\dots,z_n)$,
and $\delta=U(\tilde {\mathbf z})-U(\mathbf z)$;

\hspace{0.5cm} if $\delta<0$ then set $\mathbf z=\tilde {\mathbf z}$;

\hspace{0.5cm} if $\delta>0$ then with probability $\exp(-\delta)$ set $\mathbf z=\tilde {\mathbf z}$;

\item repeat  2.\ and start sampling when the chain has effectively
  reached equilibrium.
\end{enumerate}

\subsection{Inference methods}\label{s:methods}

First, assume that we observe both $X_W=x_W$ and $Y_W=y_W$, with
$x_W\subseteq y_W$, and let $\bar x_W=y_W\setminus x_W$. Fitting a
parametric model for $Y$ in this setting is a standard problem of
spatial statistics; see \cite{LMR15} if $Y$ is a determinantal point process;
or \cite{illian:penttinen:stoyan:stoyan:08} 
if $Y$ is a Mat{\'e}rn hard core point process of type I or II. For estimation
of parameters related to $\Pi$, assume that
$M_0(r)=M_0(r|q,\theta_\Pi)$ apart from $q$ depends 
on  a parameter $\theta_\Pi$. 
A natural idea is to base the estimation on the conditional distribution of $X_W$ given $Y_W=y_W$, which
has probability mass function
\begin{equation}\label{e:pmd}
p(x_W|y_W)=\mathrm E\left[\left\{\prod_{u\in x_W}\Pi(u)\right\}
 \left\{\prod_{v\in y_W\setminus x_W} (1-\Pi(v))\right\}\right].
\end{equation}
Since \eqref{e:pmd}
 is in general intractable, we consider instead 
  composite likelihoods for marginal distributions 
of $X_W$ given $Y_W=y_W$, noticing that conditional
on $Y_W=y_W$, 
\begin{itemize}
\item a point $u\in y_W$ is in $x_W$ with probability 
$\mathrm E[\Pi(u)]=q$, and
in $\bar x_W$ with probability $1-q$, 
\item for a pair of distinct
points $\{u,v\}\subseteq y_W$, 
\begin{itemize}
\item 
$\{u,v\}\subseteq x_W$ with
probability $\mathrm E\left\{
    \Pi(u)\Pi(v)\right\}=q^2 M_0(\|v-u\||q,\theta_\Pi)$, 
\item $\{u,v\}\subseteq\bar x_W$ with
probability $\mathrm E\left\{(1-\Pi(u))(1-\Pi(v))\right\}=1-2q
+q^2  M_0(\|v-u\||q,\theta_\Pi)$, 
\item $u\in x_W$ and $v\in\bar x_W$ with
probability $\mathrm E\left\{ \Pi(u)(1-\Pi(v))\right\}=q-q^2 M_0
(\|v-u\||q,\theta_\Pi)$. 
\end{itemize}
\end{itemize}
Conditional
on $Y_W=y_W$,   we 
define the first order
 composite likelihood $CL_1(x_W|y_W;q)$ as the product of the marginal
 selection/deletion probabilities for each of the points in $y_W$, i.e. 
\begin{align}\label{CL1}
CL_1(x_W|y_W;q)=
q^{n(x_W)} (1-q)^{n(\bar x_W)},
\end{align}
and the second order composite likelihood $CL_2(x_W|y_W;q,\theta_\Pi)$
by the product over all
unordered pairs of points in $y_W$, considering the probability
whether those points have been retained or deleted, i.e.\ 
\begin{multline}\label{CL2}
CL_2(x_W|y_W;q,\theta_\Pi)=
\left[ \prod_{\{u,v\}\subseteq x_W} q^2 M_0(\|v-u\||q,\theta_\Pi)\right]\\
\left[ \prod_{\{u,v\}\subseteq \bar x_W} 
\left\{1-2q +q^2  M_0(\|v-u\||q,\theta_\Pi)\right\} \right]
\left[\prod_{u\in x_W,v\in \bar x_W} \left\{q-q^2 M_0 (\|v-u\||q,\theta_\Pi)\right\}\right].
\end{multline}
Maximizing \eqref{CL1} yields the natural estimate 
\begin{equation}\label{q_est} \hat q = n(x_W)/n(y_W).\end{equation}
Inserting this into \eqref{CL2}, the maximization of $CL_2(x_W|y_W;\hat q,\theta_\Pi)$ then provides an estimate for the remaining parameter $\theta_\Pi$.

Second, assume that we observe $X_W=x_W$ and we want to fit a parametric model for $Y$ and $\Pi$ based on this observation. The likelihood of $X_W$ is given by the mean value  of the conditional
density \eqref{density} with respect to the distribution of
$(\Pi,Y)$ on $W$. This mean value makes likelihood inference
infeasible unless we use elaborate Monte Carlo procedures. 
Instead we consider estimation based on the intensity and pair correlation
function for $X$. Here one possibility is composite likelihoods (see 
\cite{moeller:waagepetersen:07} and the references therein)
and another is minimum contrast estimation procedures. Below we
concentrate on the latter.

Assume that $g_{Y,0}(r)=g_{Y,0}(r|\theta_Y)$ depends on a parameter $\theta_Y$ and as before $M_0(r)=M_0(r|q,\theta_\Pi)$
depends on $q$ and  $\theta_\Pi$. 
A natural and unbiased estimate of the intensity $\rho_X$ is
$\hat\rho_X=n(x_W)/|W|$, i.e.\ the observed number of points divided by the
Lebesgue measure of $W$. Given an estimate $\hat q$ of $q$,
the relation \eqref{e:rhonrhon} provides the estimate
$\hat\rho_Y = \hat q \hat\rho_X$ of $\rho_Y$. The estimation 
problem thereby reduces to estimating $(q,\theta_Y,\theta_\Pi)$. 
By  \eqref{e:rhonrhon},  this can be achieved by minimum contrast estimation based on the pair correlation function of $X$: 
\begin{equation}\label{contrast_g} (\hat q, \hat \theta_Y, \hat \theta_\Pi)=\underset{q,\theta_Y,\theta_\Pi}{\textrm{argmin}} \int_{r_{l}}^{r_{u}} \left\{M_0(r|q,\theta_\Pi)^c g_{Y,0}(r|\theta_Y)^c - \hat g_{X,0}(r)^c\right\}^2 {\rm d} r \end{equation}
where $0\leq r_l<r_u$ and $c>0$ are user-specified parameters and
$\hat g_{X,0}$ is a non-parametric kernel estimate of $g_{X,0}$ based
on the data $x_W$
(we use the default estimate
provided by \texttt{spatstat}). 
For a rectangular observation window $W$ with minimal side length
$\ell$, we chose
after some experimentation, 
$c=1$, $r_l=\ell/100$ and $r_u=\ell/4$.

Alternatively, Ripley's $K$-function can be used instead of the
pair correlation function in \eqref{contrast_g}, where we choose 
$r_l$ and
$r_u$ as above but let $c=0.5$ (following \cite{Diggle:03}). 
For the models considered in this paper, the
theoretical $K$-function, given for $d=2$ by $K(r)=2\pi\int_0^r t
g_{X,0}(t)\rm d t$, has to be
approximated by numerical methods. 

Moreover, the minimum contrast estimates obtained from the pair correlation and the
$K$-function can be combined to provide a better estimate. 
 We refer to \cite{Lavancier:Rochet:2014} for details and
consider just the example of two estimators $\hat q_g$ and $\hat q_K$
for $q$. Then the idea is to seek the weights $(\lambda_1,\lambda_2)\in\mathbb R^2$ with $\lambda_1+\lambda_2=1$ such that the linear combination $\lambda_1 \hat q_g + \lambda_2 \hat q_K$ has a minimal mean square error. The solution is 
$(\lambda_1,\lambda_2)^\top = \Sigma^{-1} \mathbf 1 / (\mathbf 1^\top
\Sigma^{-1} \mathbf 1)$, where $\Sigma$ is the mean square error
matrix of $(\hat q_g,\hat q_K)$ and $\mathbf 1=(1,1)^\top$. An
adaptive choice is obtained by replacing $\Sigma$ by an estimate
$\hat\Sigma$ in the previous formula, where $\hat\Sigma$ can be
obtained by parametric bootstrap. This `average' approach may also be used to combine several estimates   for different values of $c$ in \eqref{contrast_g}. From our experience, this does not improve significantly on our basic choice of $c$ suggested above and we do not consider this generalization in the following.  

\subsection{Simulation study}\label{s:simu}

We carried out a simulation study for the following four models when
$d=2$ and $W$ is a unit square:
\begin{enumerate}
\item $Y$ is a  DPP with Gaussian kernel $C_0(r)= \rho_Y
  \mathrm e^{-(r/\alpha)^2}$, where $\rho_Y=1000$ and $\alpha=0.015$,
  and $-\log\Pi$ is a squared zero-mean Gaussian process given by
  \eqref{pi-gauss} with $k=1$, correlation function $R_0(r)=
  \mathrm e^{-(r/s)^2}$ where $s=0.05$, and variance $\kappa$ 
deduced from  \eqref{eq:qkappa} with $q=0.5$. 
\item  $Y$ is a  DPP as above and $\Pi$ is the characteristic function
  for the Boolean model given by \eqref{pi_bool} with deterministic
  radius $\Delta_0=0.05$ and where $\rho_\Psi$ is deduced from
  \eqref{e:h} with $q=0.5$.
\item $Y$ is  a  Mat{\'e}rn hard core process of type II
with hardcore distance $D=0.015$ and $\rho_\Phi=1736$ yielding 
 $\rho_{Y_{II}}=1000$, cf.\ \eqref{e:rhoI-II}, while $\Pi$ is as in model 1.
\item $Y$ is as in model 3., and $\Pi$ is as in model 2.
\end{enumerate}
In each case, 100 independent 
realizations of $X$ were generated on the unit square. Some examples are shown in Figure~\ref{fig:examples}. 

First, we assumed that   $X$ and $Y$ are both observed. We did not fit a parametric model for $Y$, which is a standard inference  problem as explained in  Section~\ref{s:methods},  but we estimated $q$ and $\theta_\Pi$ by the composite likelihood method detailed in the same section, where $\theta_\Pi=s$ in models
1.~and 3., and $\theta_\Pi=\Delta_0$  in models
2.~and 4.  The value $k=1$ in models
1.\ and 3.\ was assumed to be fixed.  Since  the estimation of $q$ in
this setting is easy, see \eqref{q_est}, we only report in
Table~1 some summary statistics for
$\hat\theta_\Pi$. The results demonstrate  good performances of the
maximum composite likelihood estimator.

Second, we assumed that only $X$ is observed. The hardcore distance $D$ in models 3.\ and 4.\ was then estimated by the
minimal pairwise distance observed in $X$, the value $k=1$ in models
1.\ and 3.\ was assumed to be fixed, and the other parameters were
fitted as explained in Section~\ref{s:methods}, either from the pair
correlation function, or from the $K$-function, or from an optimal linear
combination of the former and the latter.  The performances of the
estimators  are summarized in Table~2 
except for
$\hat\rho_X$ and $\hat D$ which are standard estimators. For the first
model, the estimation of $s$ from the $K$-function sometimes failed
because the optimization procedure did not find a minimum. In those
circumstances, the figures in Table~2 
marked with an
asterisk are computed from only $90\%$ of the simulated point
patterns. Overall, the estimation based on the pair correlation
function $g$ seems more reliable than the estimation based on $K$,
cf.\ Table~2. 
The average estimator ($AV$) based on an optimal linear combination
always outperforms the two previous methods in terms of the mean
square error. The weights used for the combination are reported in
Table~2.

\begin{table}[ht]
\label{tab:estXY}
\begin{center}
\resizebox{\textwidth}{!} {
\begin{tabular}{ccc|ccc|ccc|ccc}
  \hline
  \multicolumn{3}{c|}{Model 1 $\ (\hat s)$} & \multicolumn{3}{c|}{Model 2 $\ (\hat \Delta_0)$} & \multicolumn{3}{c|}{Model 3 $\ (\hat s)$} & \multicolumn{3}{c}{Model 4 $\ (\hat \Delta_0)$} \\ 
Mean & sd & MSE & Mean & sd & MSE &Mean & sd & MSE &Mean & sd & MSE \\\hline
0.05 & 0.006 & $3.69${\footnotesize$\times 10^{-5}$} & 0.05 & 0.004 & $1.59${\footnotesize$\times 10^{-5}$}  & 0.05 & 0.005 & $3.05${\footnotesize$\times 10^{-5}$} & 0.05 & 0.004 & $1.56${\footnotesize$\times 10^{-5}$} \\
   \hline
\end{tabular}
}
  \caption{Empirical means, standard deviations (sd), and mean square
    error (MSE) of maximum composite likelihood
    estimates for the parameters of the model for $\Pi$, when $X$ and $Y$ are
    observed within a unit square. The results are based on 100
    simulated datasets for each of the four models of Section~\ref{s:simu}. }
\end{center}
\end{table}

\begin{table}[ht]
\label{tab:est}
\begin{center}
\resizebox{\textwidth}{!} {
\begin{tabular}{cr|ccc|ccc|cccc}
  \hline
  & &\multicolumn{3}{c|}{$g$} & \multicolumn{3}{c|}{$K$} & \multicolumn{4}{c}{$AV$} \\ 
Model   &  & Mean & sd & MSE & Mean & sd & MSE & Mean & sd & MSE & Weight\\
 \hline
 1 & $\hat q$ & 0.48 & 0.15 & 0.023 & 0.49 & 0.21 & 0.045 & 0.48 & 0.14 & 0.021 &  (0.8,0.2) \\
  &$\hat \alpha$ & 0.016 & 0.0030 & $9.6${\footnotesize$\times 10^{-6}$} & 0.014 & 0.0039 & $15.6${\footnotesize$\times 10^{-6}$} & 0.015 & 0.0028 & $8.1${\footnotesize$\times 10^{-6}$} &  (0.7,0.3)\\
 &$\hat s$ & 0.053 & 0.018 & $3${\footnotesize$\times 10^{-4}$} & $0.059^*$ & $0.049^*$ & $24^*${\footnotesize$\times 10^{-4}$} &  0.053 & 0.018 & $3${\footnotesize$\times 10^{-4}$}  &  (1,0)\\
   \hline
 2 & $\hat q$ & 0.52 & 0.055 & 0.003 & 0.50 & 0.116 & 0.013 &0.52 & 0.055 & 0.003 &  (1,0) \\
  &$\hat \alpha$ & 0.015 & 0.0014 & $2${\footnotesize$\times 10^{-6}$} & 0.015 & 0.0038 & $14${\footnotesize$\times 10^{-6}$} &  0.015 & 0.0014 & $2${\footnotesize$\times 10^{-6}$} &  (1,0)\\
  &$\hat \Delta_{0}$ & 0.052 & 0.017 & $3${\footnotesize$\times 10^{-4}$} & 0.056 & 0.034 & $12${\footnotesize$\times 10^{-4}$}& 0.052 & 0.017 & $3${\footnotesize$\times 10^{-4}$} &  (1,0)\\
      \hline
      3 & $\hat q$ & 0.66 & 0.05 & 0.029 & 0.51 & 0.16 & 0.025 & 0.58 & 0.09 & 0.015 &  (0.4,0.6) \\
  &$\hat s$ & 0.07 & 0.04 & 0.0019 & 0.08 & 0.09 & 0.0098 &  0.07 & 0.04 & 0.0019 &  (1,0)\\
      \hline
  4 & $\hat q$ & 0.56 & 0.050 & 0.0066 & 0.50 & 0.077 & 0.0058 & 0.53 & 0.061 & 0.0045 &  (0.5,0.5) \\
  &$\hat \Delta_{0}$ & 0.058 & 0.009 & $1.5${\footnotesize$\times 10^{-4}$} & 0.052 & 0.022 & $4.9${\footnotesize$\times 10^{-4}$} &   0.058 & 0.009 & $1.5${\footnotesize$\times 10^{-4}$} &  (1,0)\\
      \hline

\end{tabular}
} 
\caption{Empirical means, standard deviations (sd), and mean square error (MSE) of
    parameter estimates based on 100 simulated datasets for the four
    models of Section~\ref{s:simu}, when only $X$ is observed within a
    unit square and different minimum contrast estimation procedures are used.}
\end{center}
\end{table}

\section{Data examples}\label{s:applications}

This section illustrates how our statistical methodology applies for two real
datasets when $Y$ is observed (Section~\ref{s:allogny}) or not (Section~\ref{s:ponderosa}).

\subsection{Allogny dataset}\label{s:allogny}

\begin{figure}[h]
\begin{center}
\includegraphics[angle=0,scale=0.15]{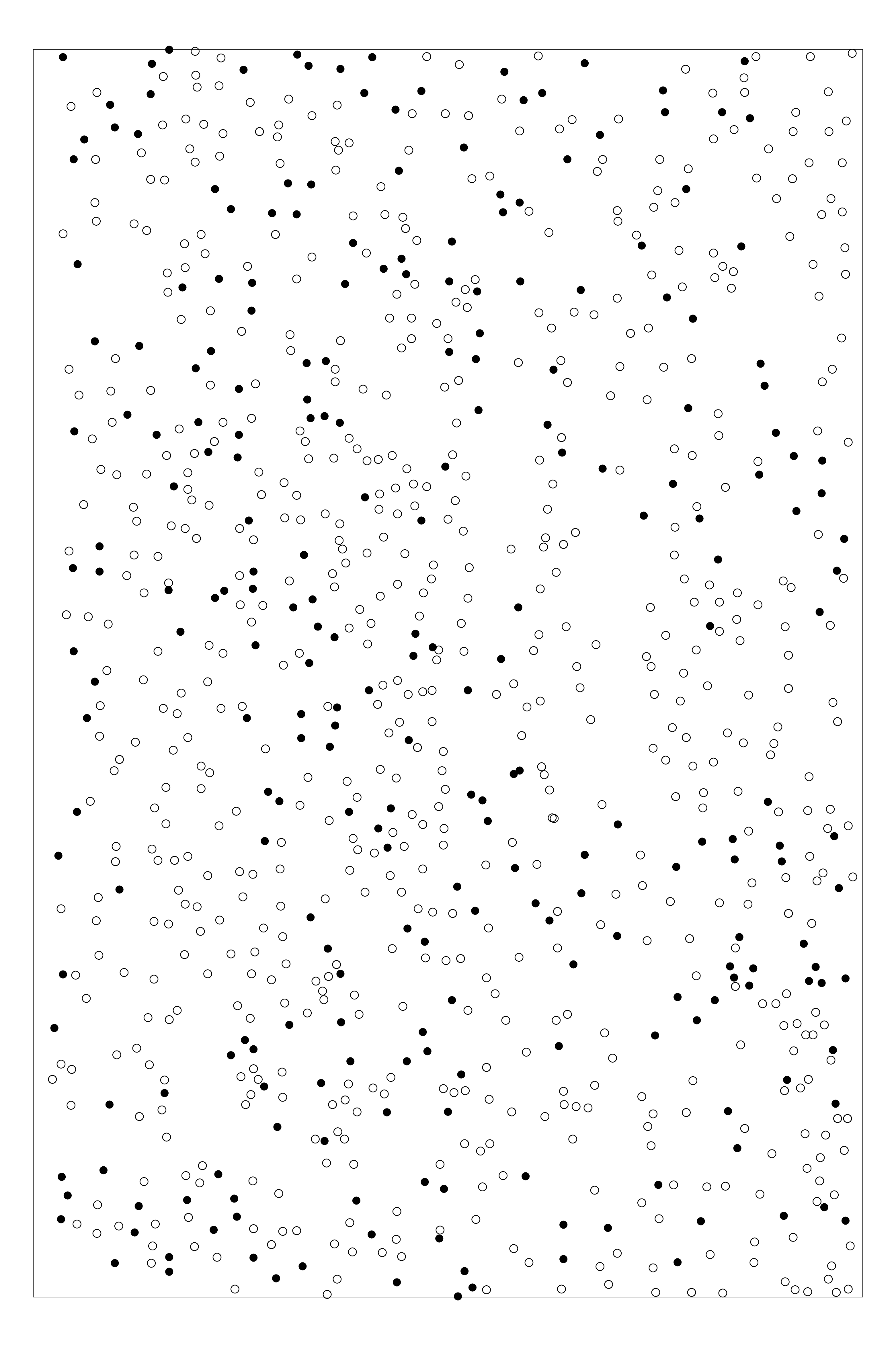} \includegraphics[angle=0,scale=0.15]{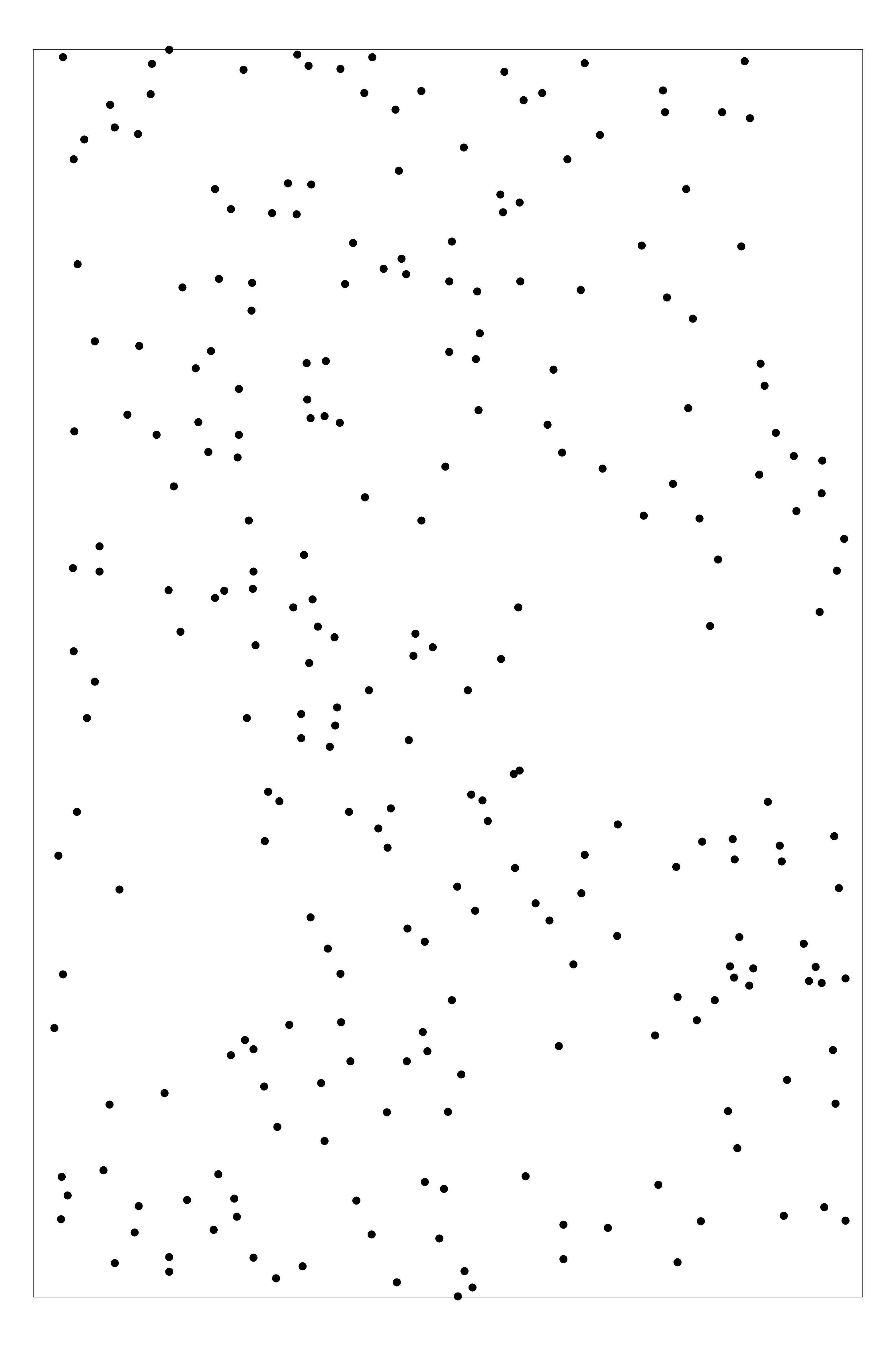}   \includegraphics[scale=0.22]{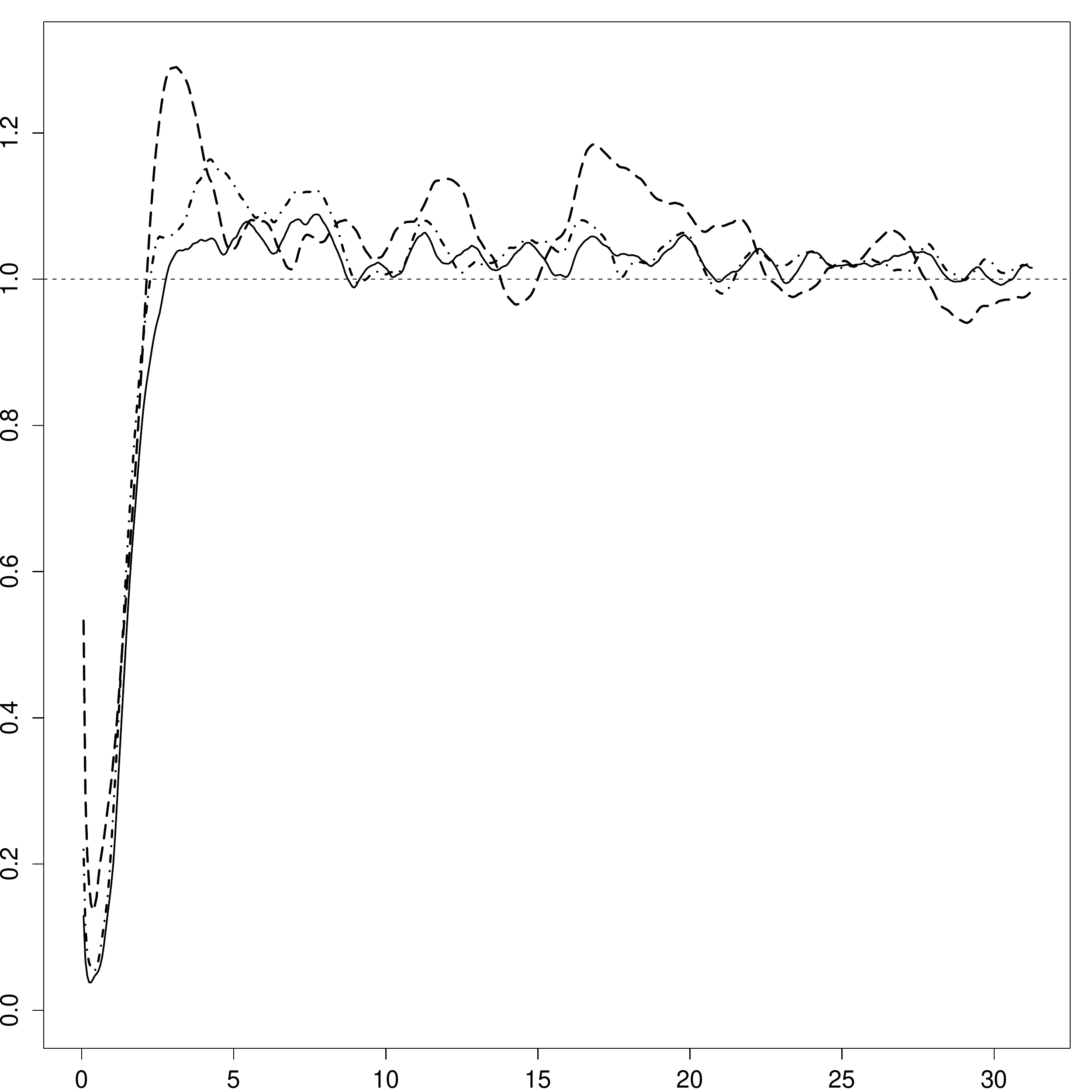} 
\caption{Left: Allogny dataset of sound oaks (circles) and splited
  oaks (solid points). Middle: splited oaks only. Right:
  non-parametric estimate of the  pair correlation function for all trees (solid curve), the sound oaks (dotted-dashed curve), and the splited oaks (dashed curve).}\label{allogny}
\end{center}
\end{figure}

Figure~\ref{allogny} shows the position of 910 oak trees in a
$125\times 188$m region at Allogny, France, where the
256 solid points correspond to
"splited oaks",  damaged by frost shake, and  the 654 remaining trees
("sound oaks") are represented by small circles. This dataset is
available in the \texttt{ads} library \citep{Pelissier2015}. It has
been analyzed in \cite{goreaud2003} and in
\cite{illian:penttinen:stoyan:stoyan:08}, where the question was to
decide whether frost shake is a clustered phenomenon, as the empirical
pair correlation function of the splited oaks in Figure~\ref{allogny}
may suggest. To the best of our knowledge, 
a parametric model for the dataset has yet not been
proposed and analyzed.

We apply our model to this dataset where $X$ represents the splited
oaks and $Y$ is the unmarked point pattern composed of the splited  and the sound oaks. In this application, the inclusion probabilities given by  $\Pi$ have a natural interpretation in terms of unobserved environmental conditions that locally favor frost shake. 
Specifically, following the procedure explained in  Section~\ref{s:methods},  we fit a parametric model to $\Pi$ by the composite likelihood method. We are in particularly interested here by the conditional simulation of $\Pi$ given the observation of the sound oaks and the splited oaks. 

\begin{figure}[h]
\begin{center}
\includegraphics[scale=0.145]{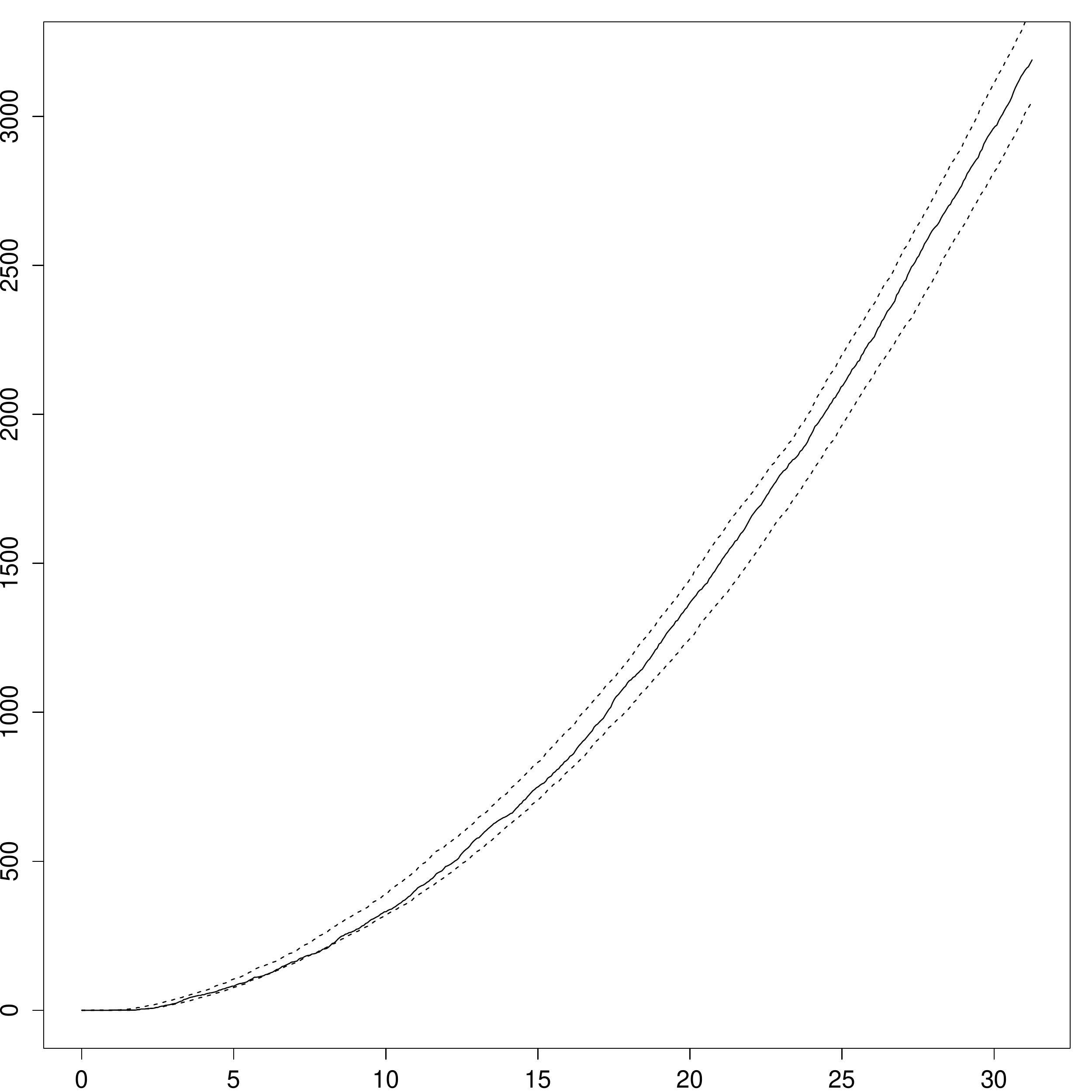}   \includegraphics[scale=0.145]{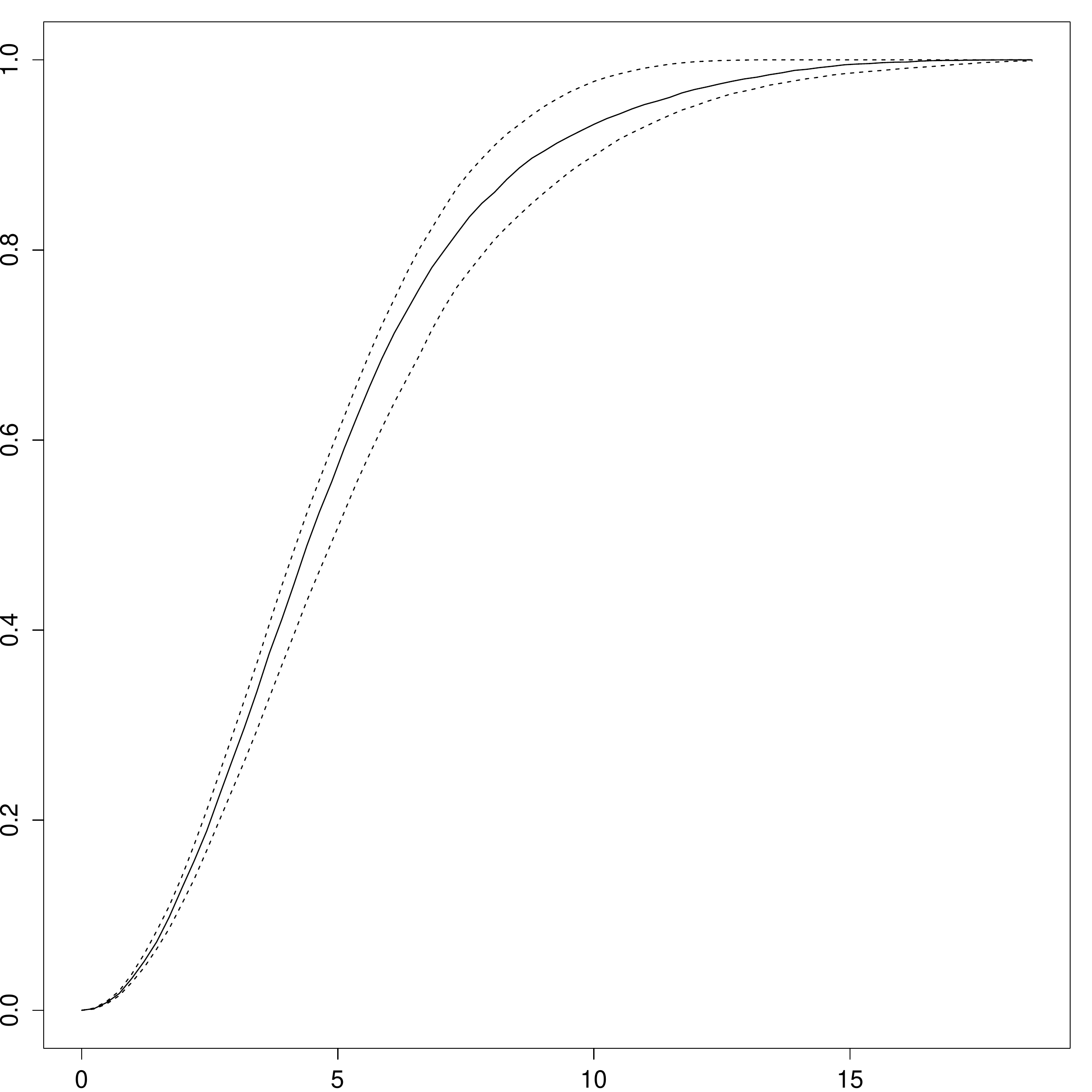} 
 \includegraphics[scale=0.145]{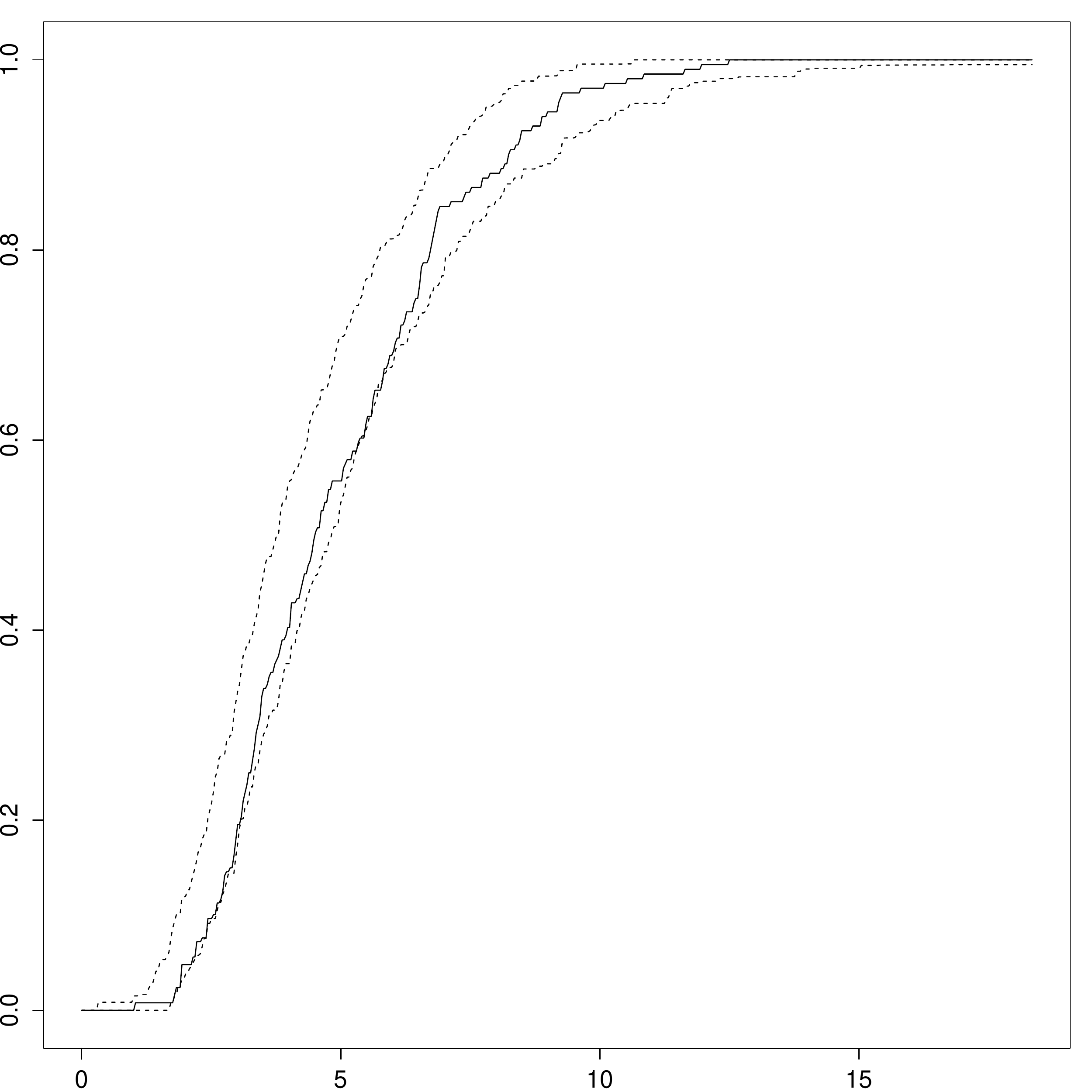}   \includegraphics[scale=0.145]{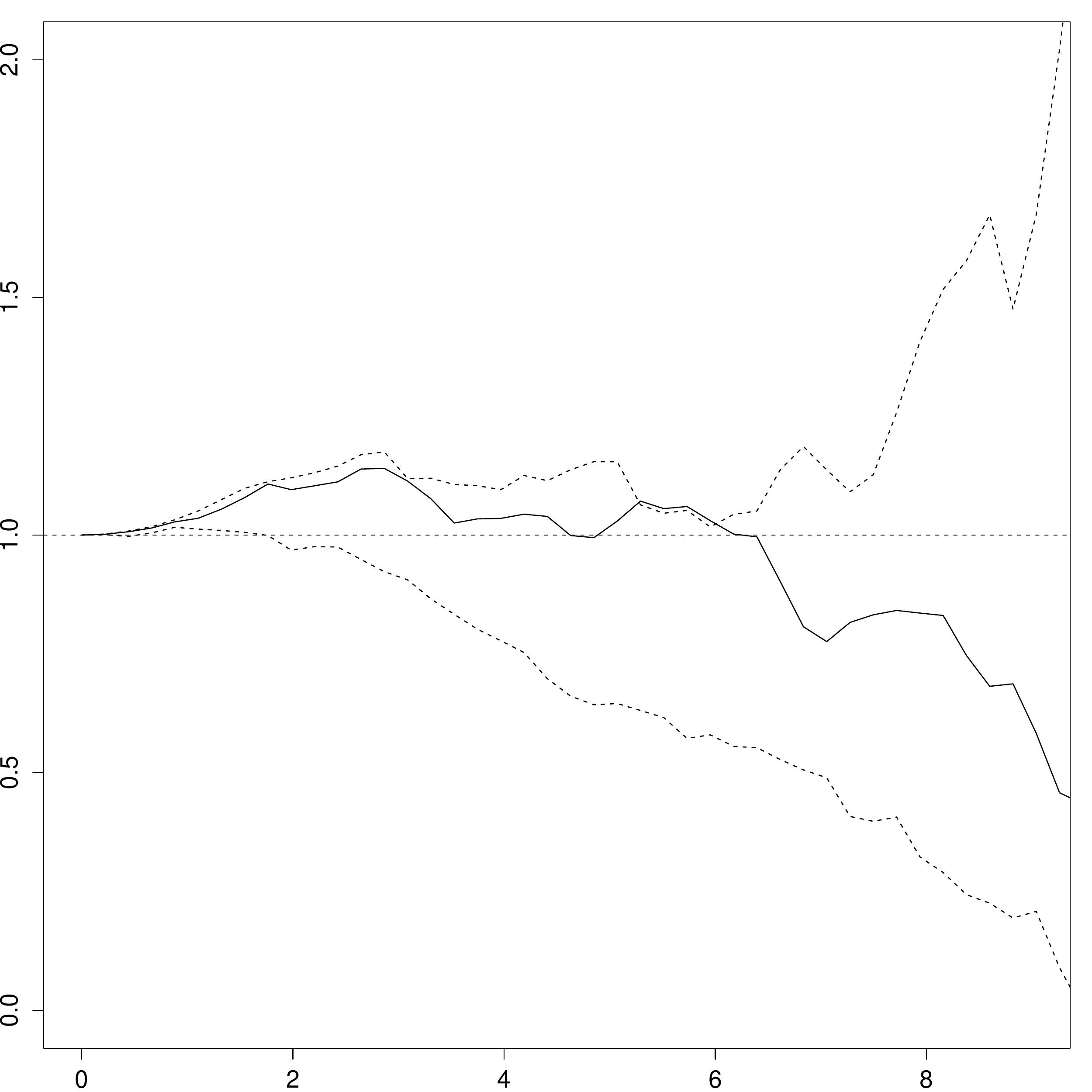} 
\caption{From left to right non-parametric estimates of the $K$, $F$,
  $G$, and $J$ functions for the splited oaks (solid lines)
  along with simulated $95\%$ pointwise envelopes obtained under the
  fitted model of Section~\ref{s:allogny} (dashed lines).}\label{fig:valid_allogny}
\end{center}
\end{figure}

\begin{figure}[h]
\begin{center}
\includegraphics[scale=0.145]{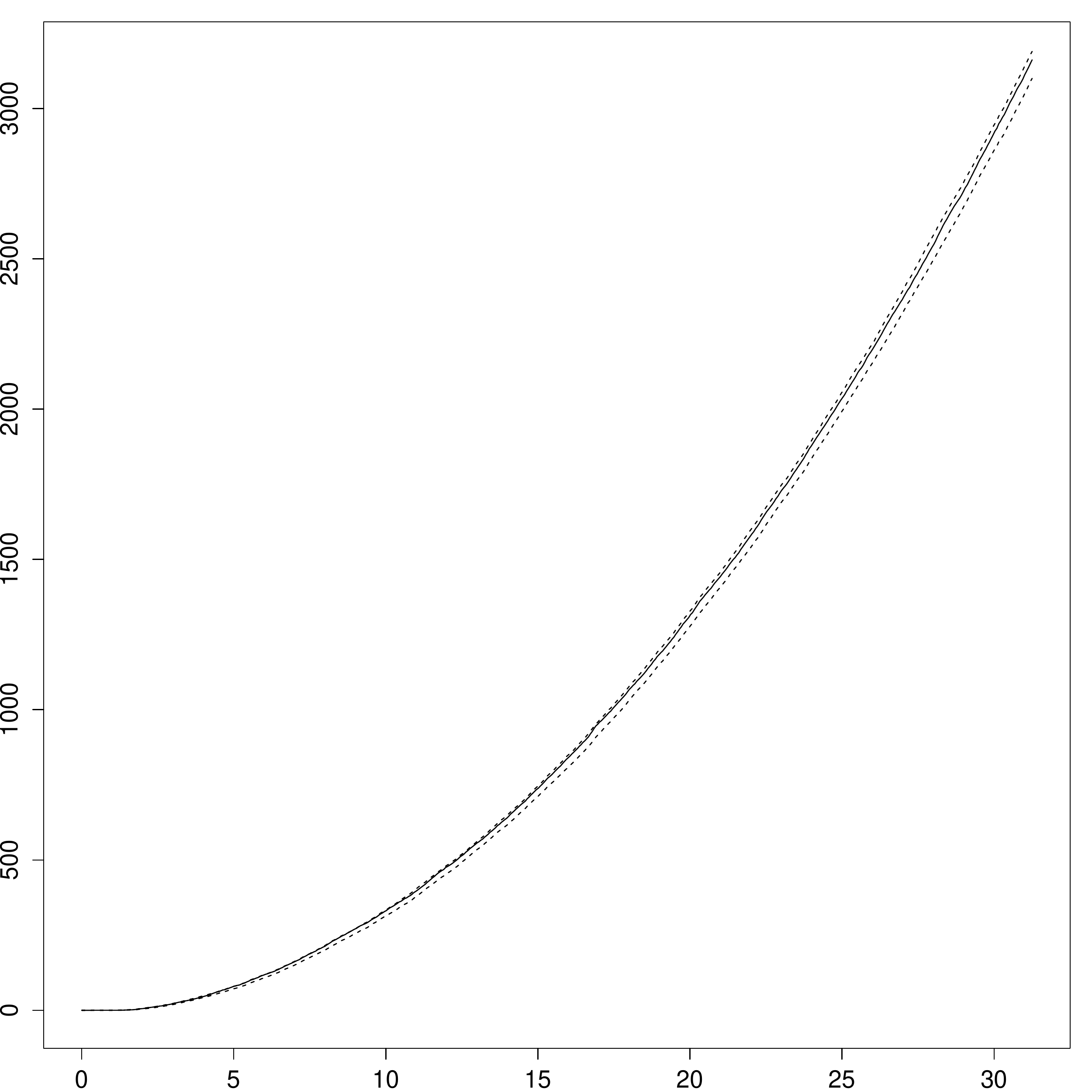}   \includegraphics[scale=0.145]{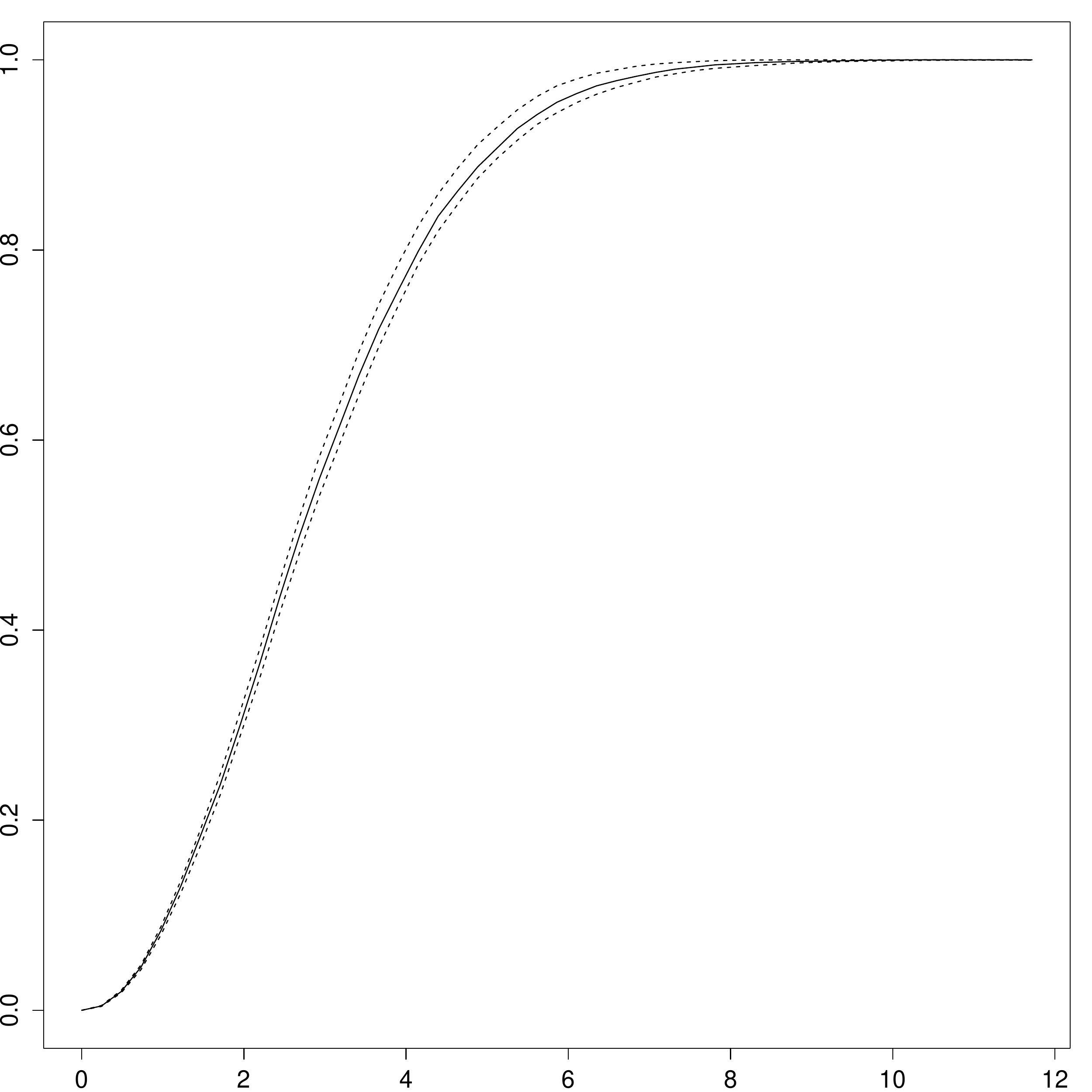} 
 \includegraphics[scale=0.145]{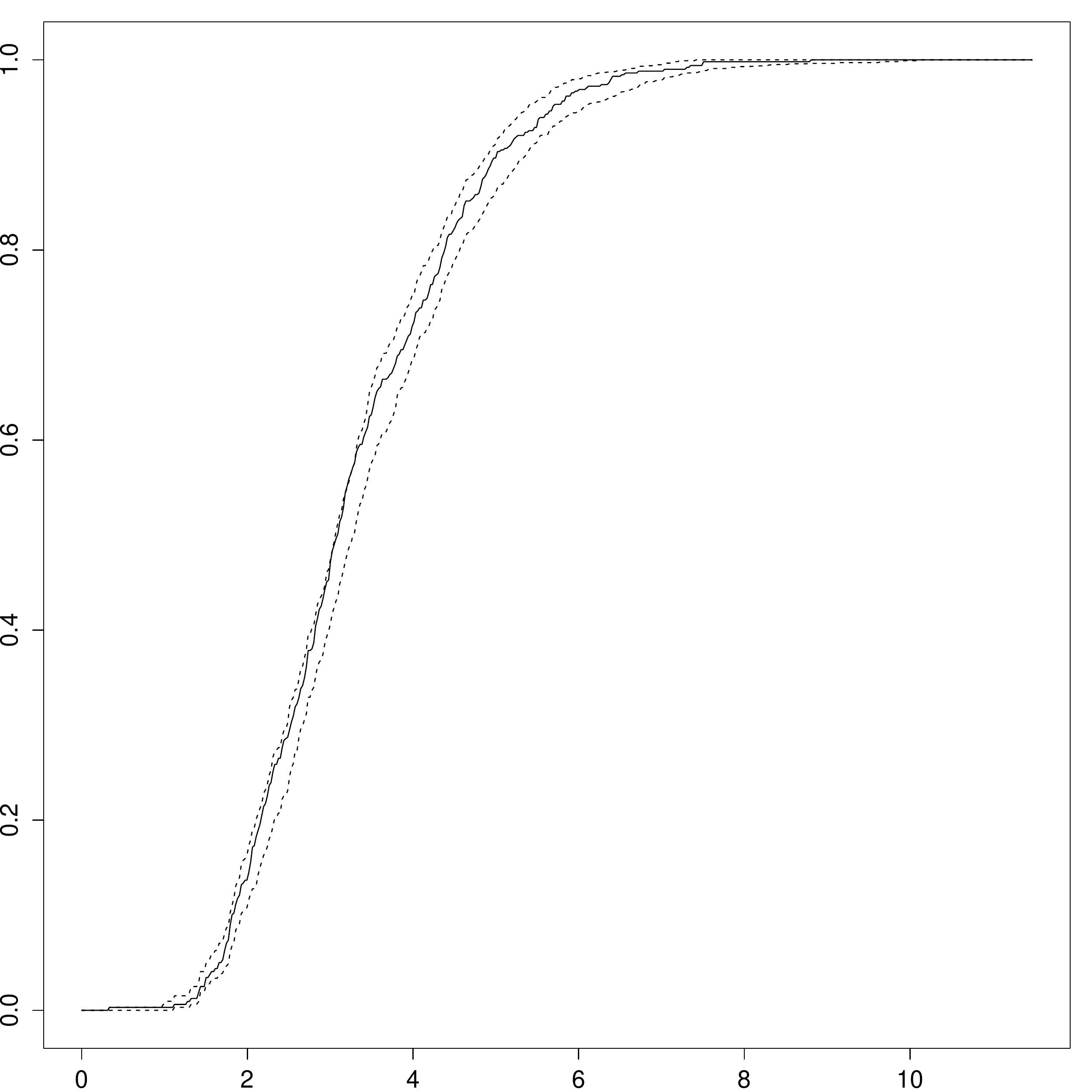}   \includegraphics[scale=0.145]{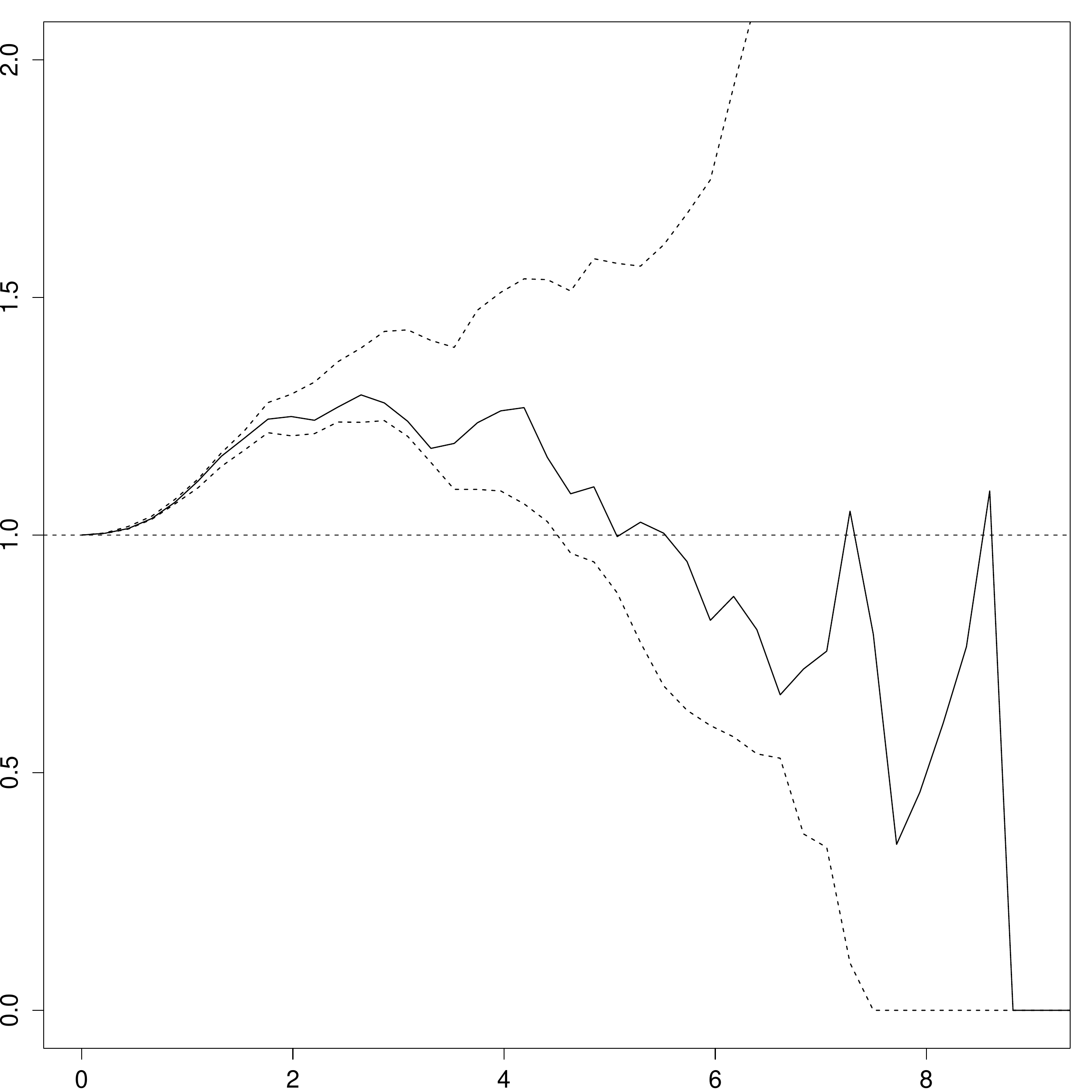} 
\caption{From left to right non-parametric estimates of the $K$, $F$,
  $G$, and $J$ functions for the sound oaks (solid lines) along
  with simulated $95\%$ pointwise envelopes  obtained under the fitted model of Section~\ref{s:allogny} (dashed lines).}\label{fig:valid_allognycompl}
\end{center}
\end{figure}

Both models presented in Sections~\ref{s:3.1}-\ref{s:3.2} can be considered for $\Pi$. 
However we think that  a  Boolean model is too simple to explain the
clustering behaviour of splited trees and we 
therefore assume that $-\log\Pi$ is a squared stationary and isotropic
Gaussian process given by \eqref{pi-gauss} with $k=1$ and $R_0$ being
a Whittle-Mat\'ern correlation function with shape parameter $\nu>0$ and
scale parameter $s>0$ (see \cite{LMR15}). The estimate of $q$ is
$\hat q=256/910=0.28$ whereby \eqref{eq:qkappa} gives
$\hat\kappa=11.8$, and by maximizing \eqref{CL2} using a grid of
values for $\nu$ and substituting $q$ by $\hat q$, we obtain $\hat
s=7.8$ and $\hat\nu=0.5$, in which case $R_0$ becomes the exponential correlation function. 
The goodness of fit is assessed by comparing
non-parametric estimates of the $K$, $F$, $G$, and $J$ functions (see
e.g.\ \cite{moeller:waagepetersen:00})
for the splited oaks with  $95\%$ pointwise envelopes of the same functions obtained  from simulations of the fitted model.
Here, in accordance with our inference procedure, the simulation of
new splited oaks is done conditionally on the tree locations, meaning that only $\Pi$ is simulated. The results are reported in Figure~\ref{fig:valid_allogny}. 
Furthermore, the same comparison is done  for the sound oaks in
Figure~\ref{fig:valid_allognycompl}. Figures~\ref{fig:valid_allogny}-\ref{fig:valid_allognycompl}
show that the goodness of fit is acceptable. 

\begin{figure}[h]
\begin{center}
\includegraphics[angle=0,scale=0.15]{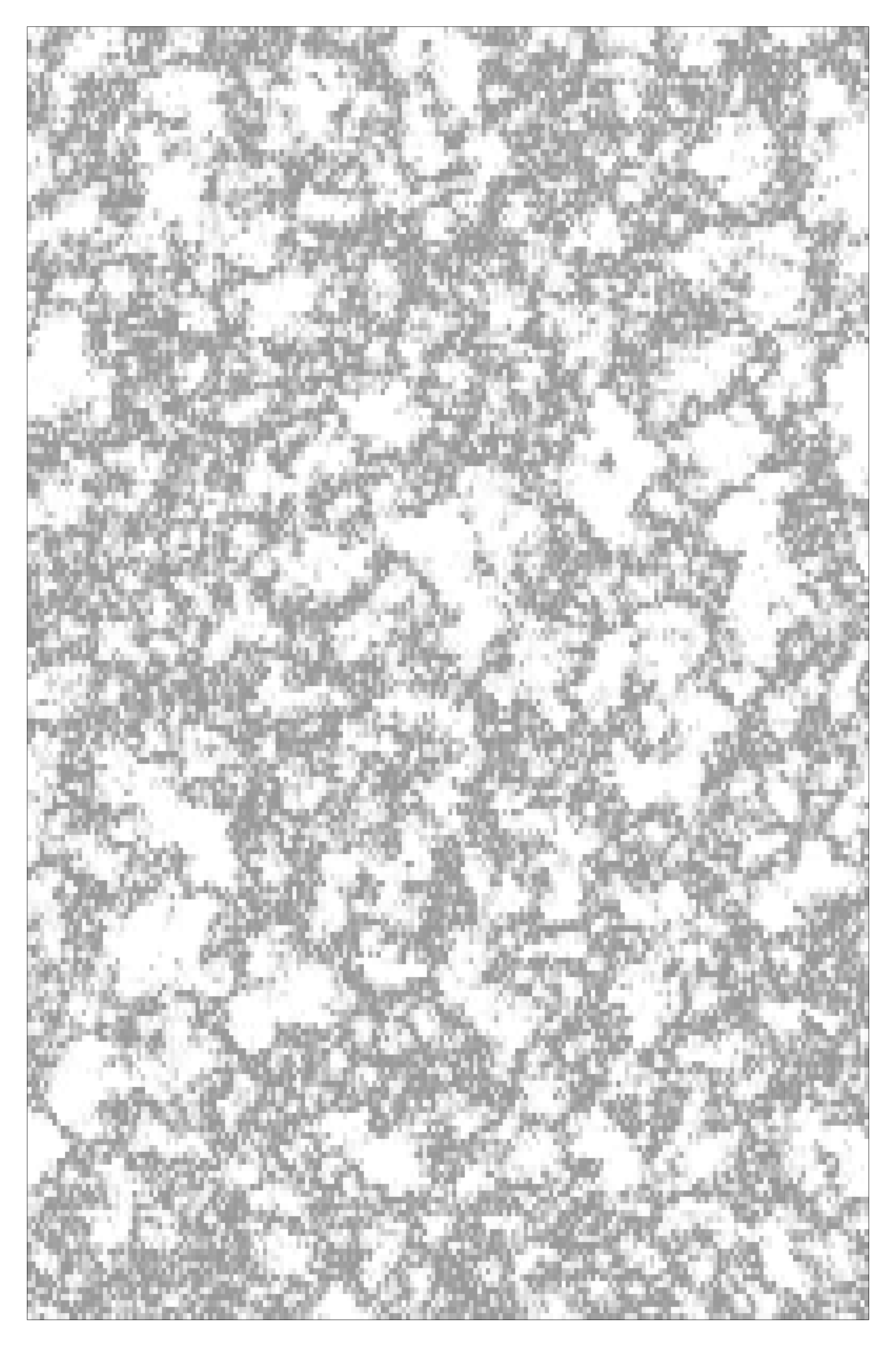} \includegraphics[angle=0,scale=0.15]{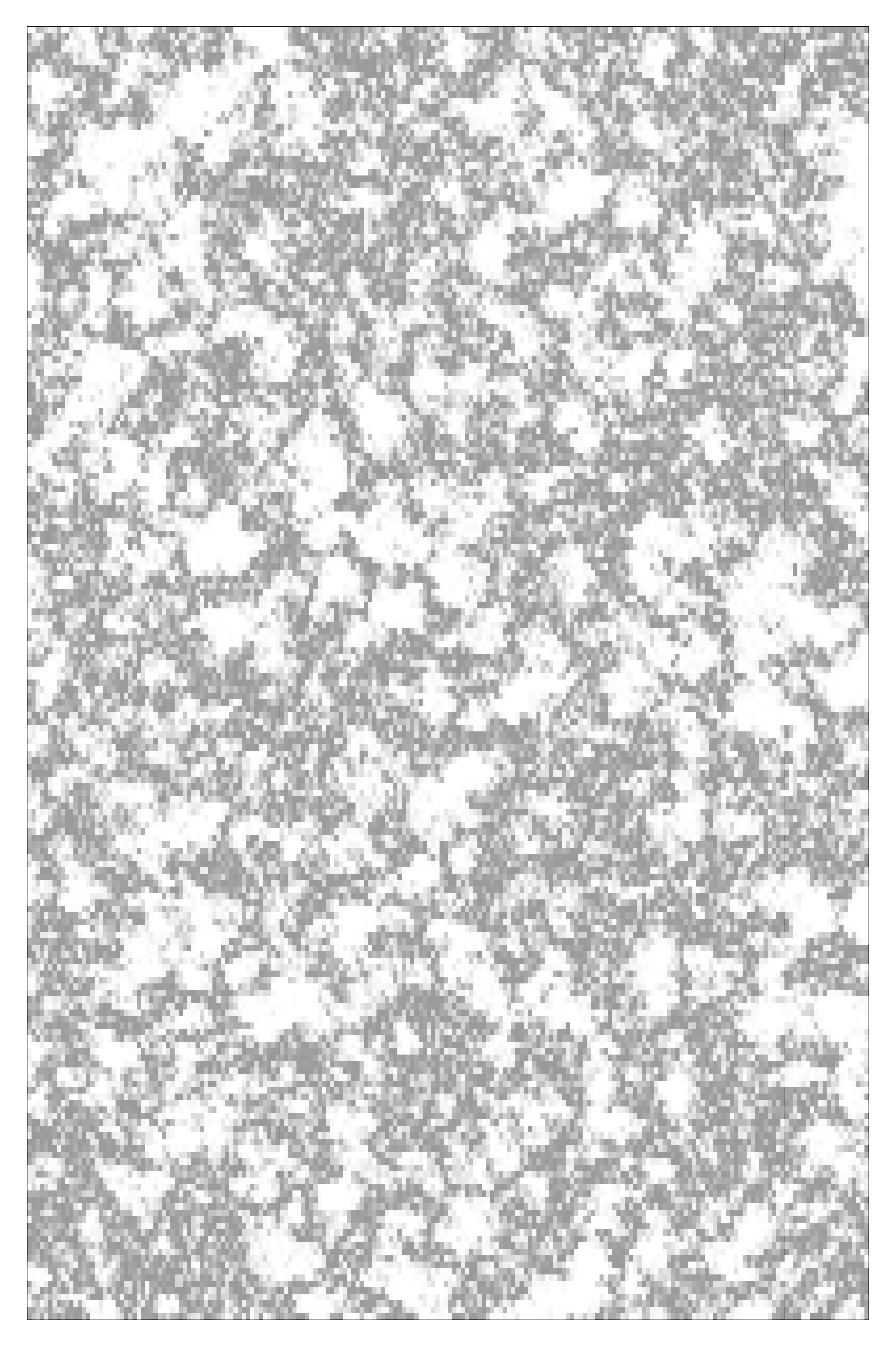}   
\includegraphics[scale=0.15]{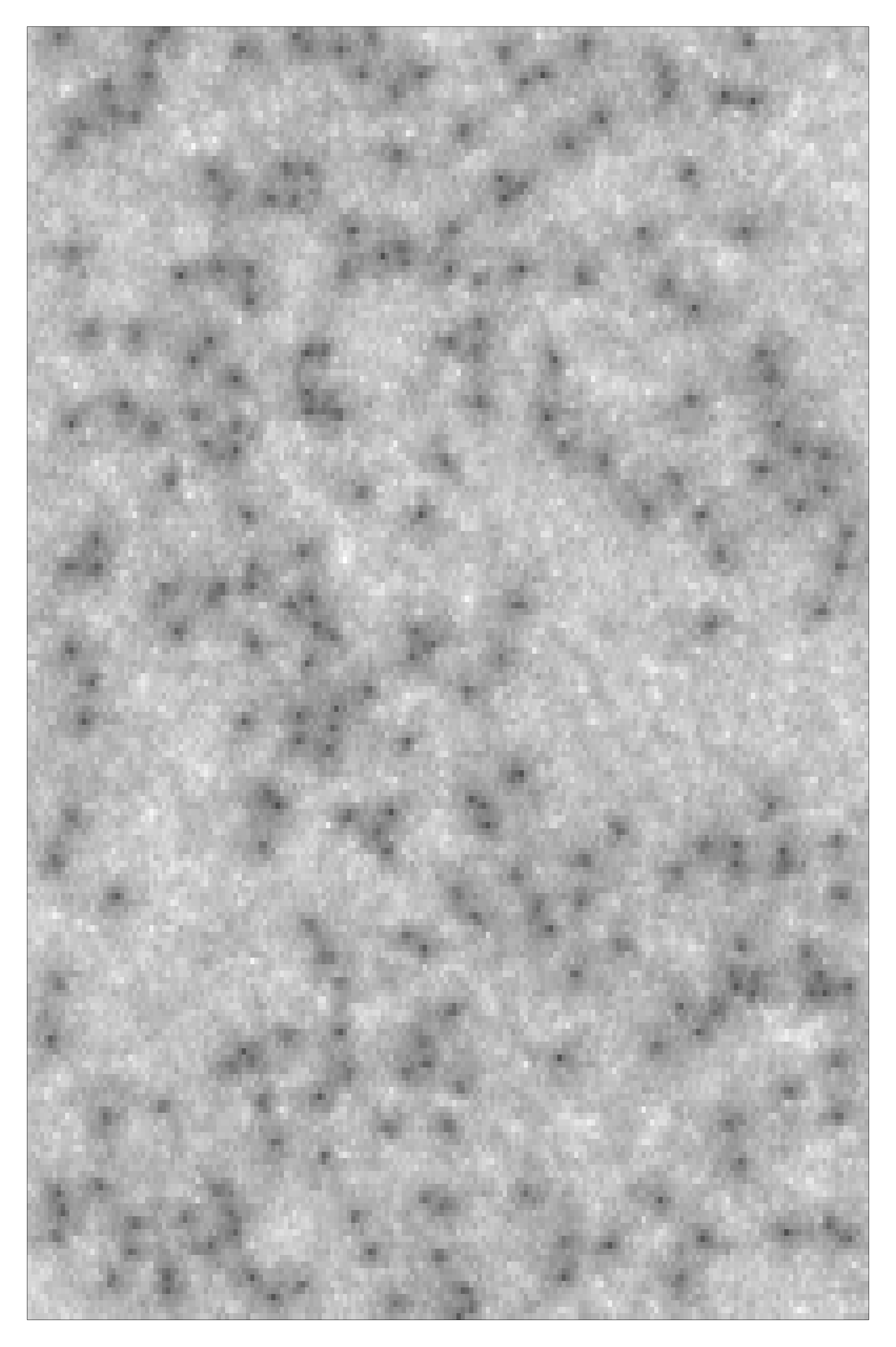}
\caption{Typical simulations (left and middle) and approximated expectation (right) of $\Pi$ under the fitted model of
  Section~\ref{s:allogny}, conditional on the splited oaks and the sound oaks.}\label{allognycond}
\end{center}
\end{figure}

Next, assuming $\Pi$ follows the fitted model, we simulate $\Pi$
conditional on the splited oaks and the sound oaks, using the two
steps procedure detailed in Section~\ref{s:simulation}. 
Figure~\ref{allognycond} shows two such realizations of $\Pi$ and an approximation of  the conditional expectation obtained from the average over 100 independent  realizations of $\Pi$. The white and lighter areas in these grayscale plots correspond to regions where frost shake seems unlikely to happen. 
 The two simulated realizations illustrate the `roughness' of $\Pi$ 
 due to the underlying exponential covariance
function. As expected, the  conditional expectation of $\Pi$ is large in the neighborhoods of splited oaks. 

\subsection{Ponderosa dataset}\label{s:ponderosa}

\begin{figure}[h]
\begin{center}
\begin{tabular}{cc} 
\includegraphics[scale=0.22]{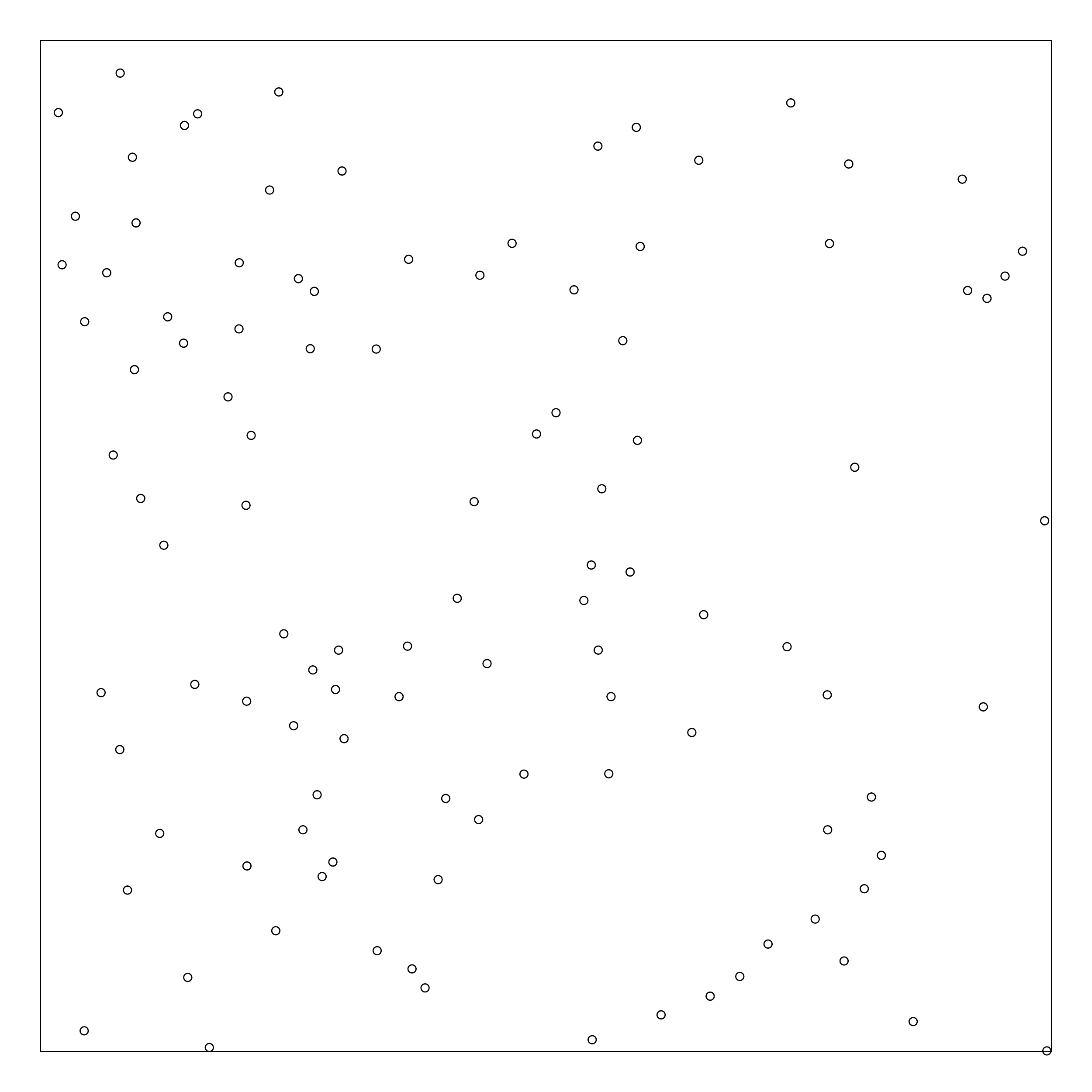}  & \includegraphics[scale=0.22]{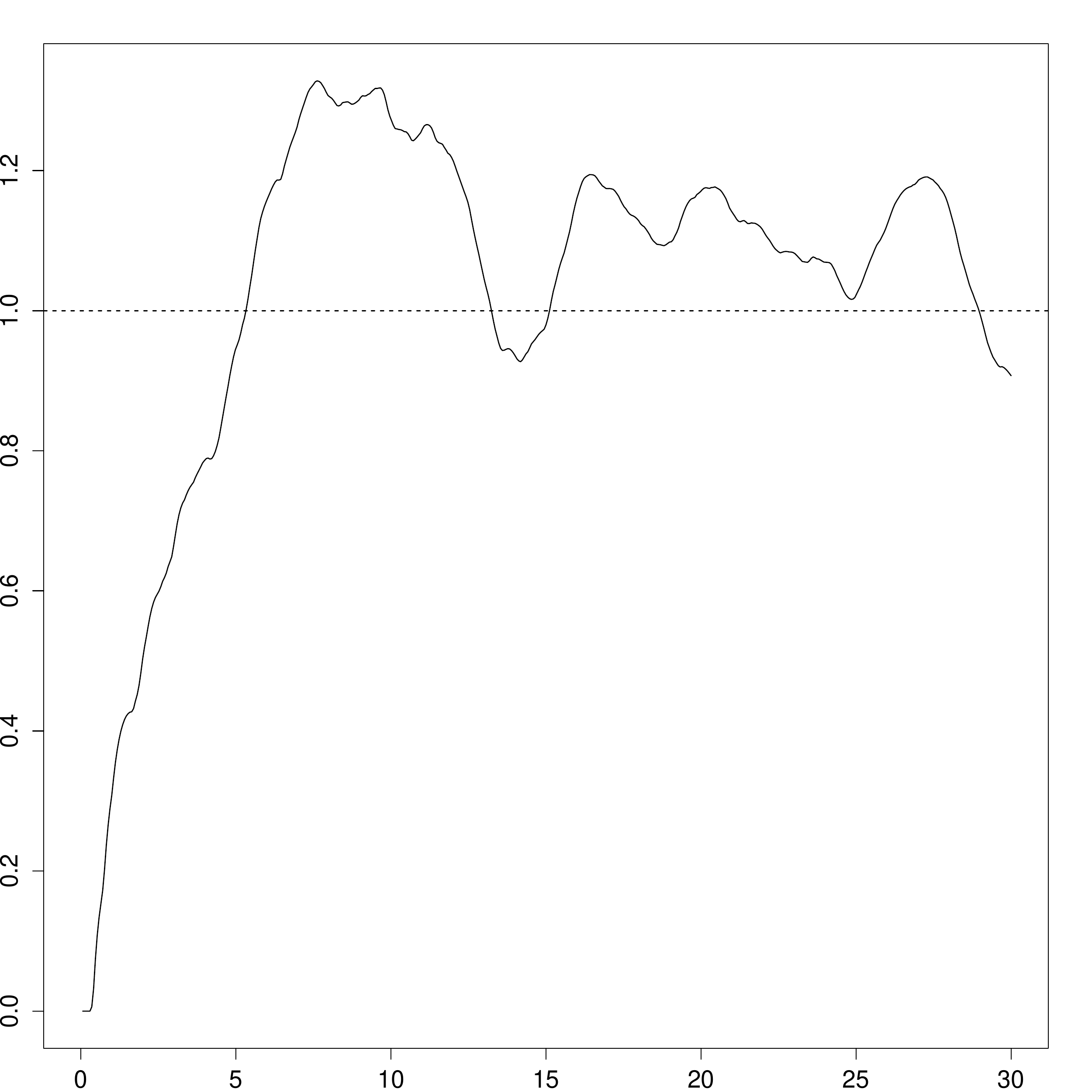} 
\end{tabular}
\caption{Left: Ponderosa dataset. Right: Non-parametric estimate of its pair correlation function.}\label{ponderosa}
\end{center}
\end{figure}

Figure~\ref{ponderosa} shows the location of 108 ponderosa pine trees in a $120\times 120$m area of  the Klamath National Forest in Northern California. This dataset is available in the \texttt{spatstat} library and was studied in \cite{getis1987}.
By a descriptive second-order analysis, the authors identified different types of  clustering between the trees, depending on the scale.  In particular they noticed that ``there are clusters of points and an apparent inhibition effect''.

We fit our model to this dataset where $Y$ is assumed to be a DPP and
$\Pi$ to follow \eqref{pi-gauss} with a Gaussian covariance function
or \eqref{pi_bool} where $\Delta_0$ is deterministic.  Three parametric
families of kernels were considered for the DPP: the Gaussian
covariance functions, the Whittle-Mat{\'e}rn covariance functions (see
\cite{LMR15}), and the Bessel-type covariance functions (see
\cite{biscio15}). These covariance functions depend on the intensity
$\rho_Y$ (which is equal to the variance), on a scale parameter
$\alpha$ and for the Whittle-Mat{\'e}rn and Bessel-type covariance
functions  on an extra shape parameter, denoted $\nu$ and $\sigma$, respectively. The Gaussian kernels family is in fact a limiting case of the two others families when $\nu$, respectively $\sigma$, tends to infinity.
It is well known that the identification of both $\alpha$ and $\nu$ (respectively $\sigma$) is difficult, even when $Y$ is fully observed and not thinned by $\Pi$.  To estimate all parameters, we use a minimum contrast method  as explained in Section~\ref{s:methods}  for different values of $\nu$  (respectively $\sigma$) on a grid and then choose the parameters giving the minimal value of the contrast function. 

\begin{figure}[h]
\begin{center}
\includegraphics[scale=0.145]{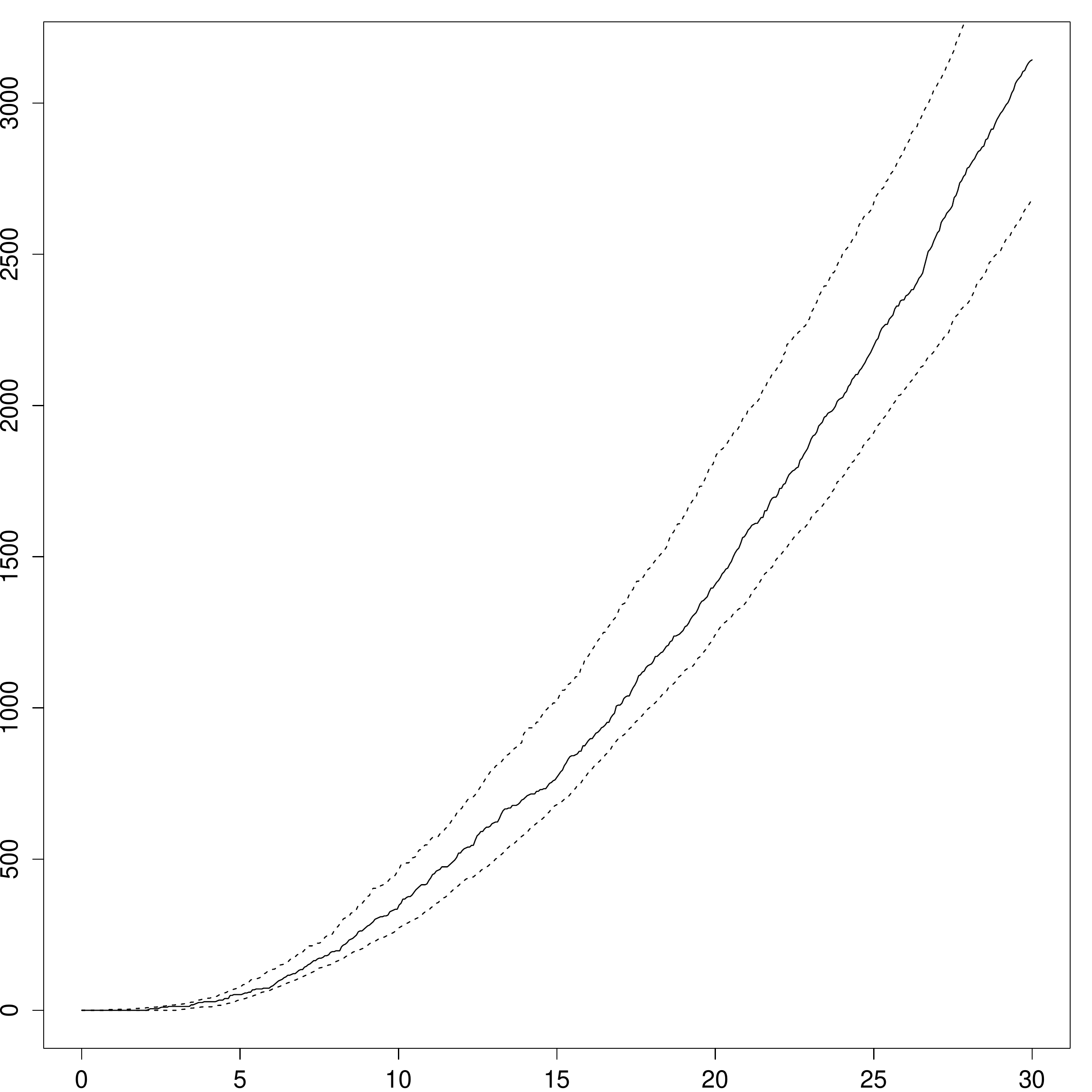}   \includegraphics[scale=0.145]{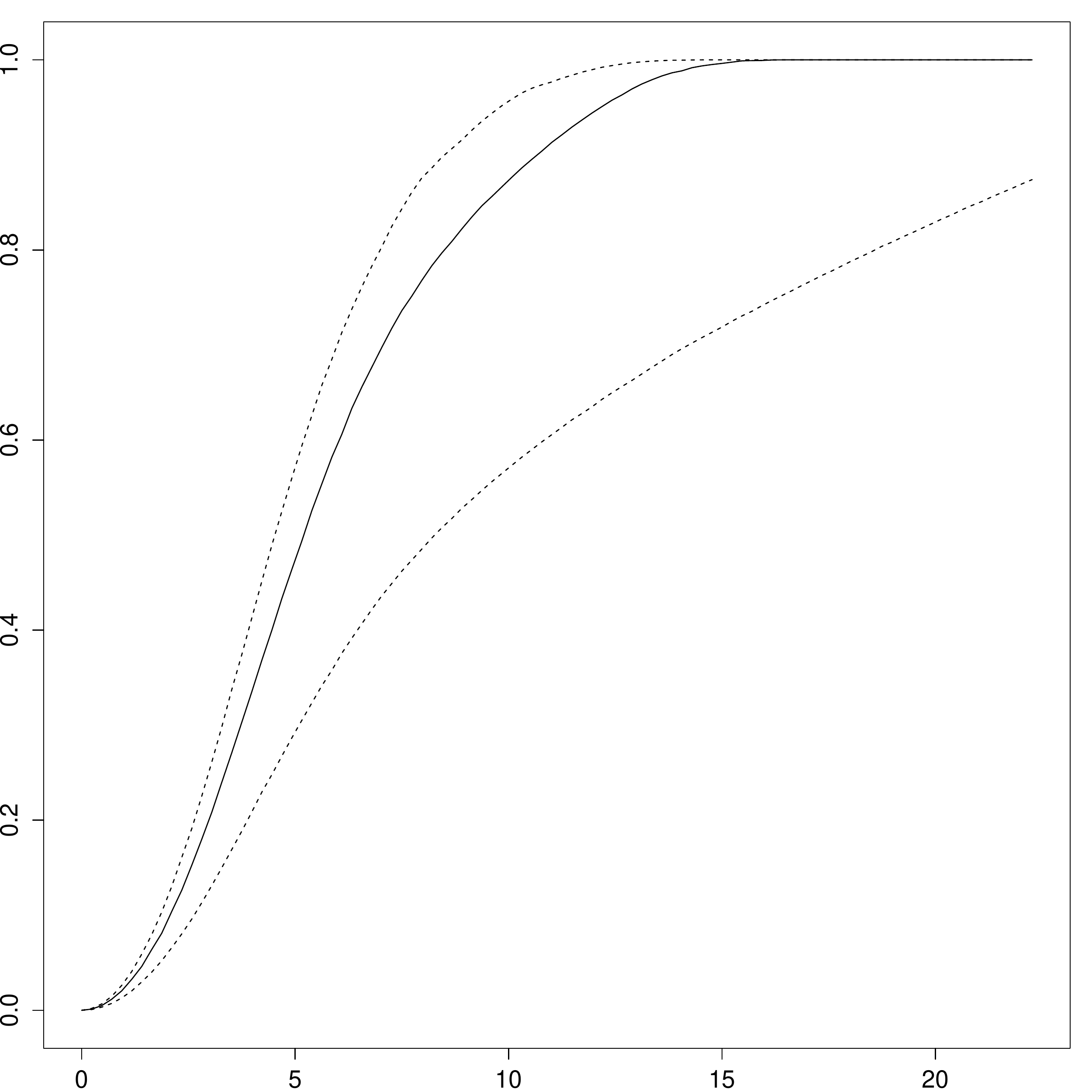} 
 \includegraphics[scale=0.145]{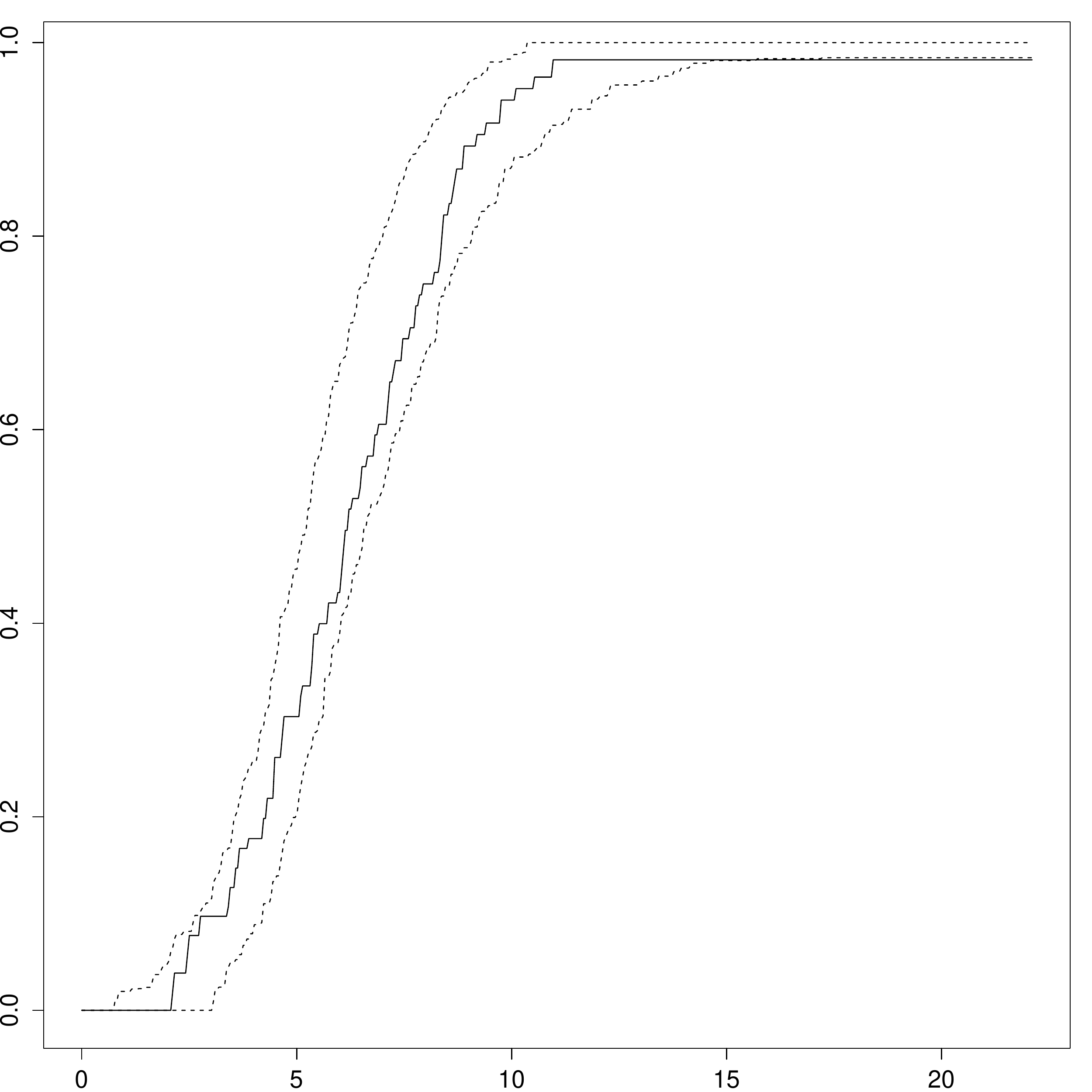}   \includegraphics[scale=0.145]{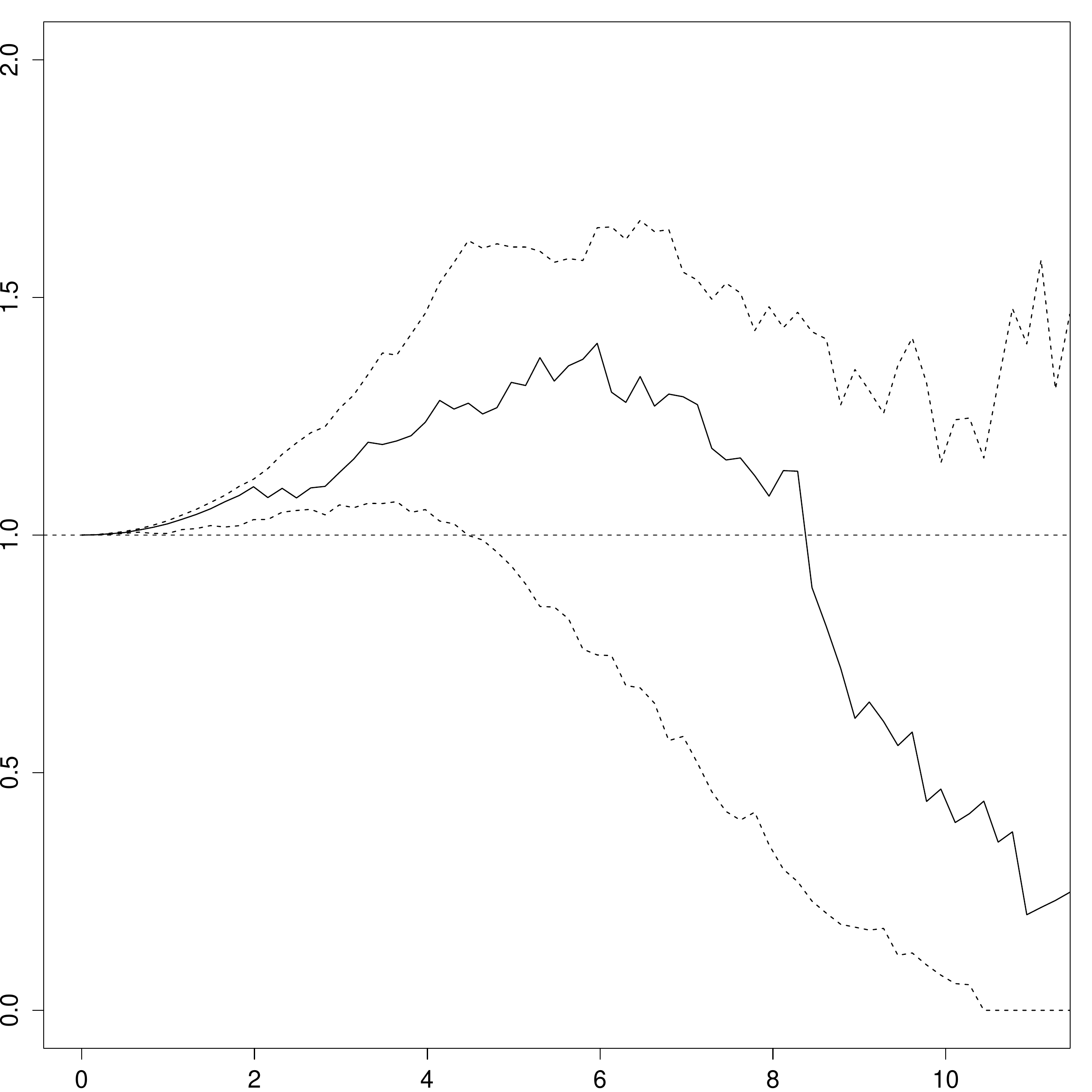} 
\caption{From left to right non-parametric estimation of the $K$, $F$,
  $G$, and $J$ functions for the Ponderosa dataset (solid lines) along
  with simulated $95\%$ pointwise envelopes under the fitted model of Section~\ref{s:ponderosa} (dashed lines).}\label{fig:valid_ponderosa}
\end{center}
\end{figure}

Among all fitted six models, we selected the one associated to the minimal
value of the contrast function based on the pair correlation function,
cf.\ \eqref{contrast_g}. The best fit was then obtained for $Y$
being the  DPP with a Gaussian covariance function and $\Pi$ the
Boolean model, where 
the minimum contrast procedure together with 
\eqref{e:h}-\eqref{e:j} 
give the estimates  
$\hat\alpha=5.26$ (with $\alpha$ being the scale parameter of the Gaussian 
covariance function), $\hat q=0.74$, and $\hat
\Delta_0=23.14$. Further, $\hat q=0.74$ together with the
natural estimate $\hat\rho_X=108/(120)^2=0.0075$ give
$\hat\rho_Y=\hat\rho_X/\hat q= 0.01$. 
When trying to improve the estimation of the selected model by the
combination method based of the $g$ and $K$ functions (described at the
end of Section~\ref{s:methods}), the best weight was (1,0), thus confirming the
choice of the contrast function based on $g$ for this model.
 The goodness of fit is assessed by comparing the
non-parametric estimates of the $K$, $F$, $G$, and $J$ functions
based on the data with  $95\%$ pointwise envelopes of the same functions obtained  from simulations of the fitted model. The result is shown in Figure~\ref{fig:valid_ponderosa} where the fitted model appears to provide a good fit. 

\subsubsection*{Acknowledgment}
Supported by the Danish Council for Independent Research | Natural
Sciences,
grant 12-124675,
"Mathematical and Statistical Analysis of Spatial Data", and by
the "Centre for Stochastic Geometry and Advanced Bioimaging",
funded by grant 8721 from the Villum Foundation.

\bibliography{bibliography}
\bibliographystyle{chicago}
\end{document}